\documentclass[prd,twocolumn,amsmath,amssymb,floatfix,superscriptaddress]{revtex4}
 \usepackage{graphicx}
 \usepackage{epstopdf}
 \usepackage{amsmath,amssymb}
 \usepackage{bm}
 \usepackage{verbatim,color}

\DeclareFontFamily{OT1}{pzc}{}
\DeclareFontShape{OT1}{pzc}{m}{it}%
             {<-> s * [1.00] pzcmi7t}{}
\DeclareMathAlphabet{\mathscr}{OT1}{pzc}%
                                 {m}{it}

\newcommand{\g}{\mathscr{g}}
\newcommand{\gbar}{\overline{\g}}

\newcommand{\be}{\begin{equation}}
\newcommand{\ee}{\end{equation}}
\newcommand{\bea}{\begin{eqnarray}}
\newcommand{\eea}{\end{eqnarray}}
\newcommand{\refeq}[1]{Eq.~(\ref{eqn:#1})}

\newcommand{\reffig}[1]{Fig.~\ref{fig:#1}}          

\newcommand{\reftab}[1]{Tab.~\ref{t:#1}}
\newcommand{\vs}{\nonumber\\}       
\newcommand{\refsec}[1]{Sec.~\ref{sec:#1}}          
\def\ba#1\ea{\begin{align}#1\end{align}}

\renewcommand{\d}{\delta}
\newcommand{\D}{\Delta}

\newcommand{\fs}[1]{\textcolor{blue}{\textbf{[FS: #1]}}}

   \newcommand{\planck}{Planck}

\begin{document}
 
   \title{Constraints on Modified Gravity from Sunyaev--Zeldovich Cluster Surveys}
 
 \author{Daisy S. Y. Mak}
 \email{suetyinm@usc.edu}
\affiliation{Physics and Astronomy Department, University of Southern California, Los Angeles, California 90089-0484, USA}

\author{Elena Pierpaoli}
 \email{pierpaol@usc.edu}
\affiliation{Physics and Astronomy Department, University of Southern California, Los Angeles, California 90089-0484, USA}

\author{Fabian Schmidt}
\email{fabians@caltech.edu}
\affiliation{Theoretical Astrophysics, California Institute of Technology, Pasadena, CA 91125, USA}

\author{Nicolo' Macellari}
 \email{pierpaol@usc.edu}
\affiliation{Physics and Astronomy Department, University of Southern California, Los Angeles, California 90089-0484, USA}

 
 \begin{abstract}
We investigate the constraining power of current and future Sunyaev-Zeldovich cluster surveys on 
the $f(R)$ gravity model.  We use a Fisher matrix approach, adopt self-calibration 
for the mass-observable scaling relation, and evaluate constraints for the 
SPT, Planck, SPTPol and ACTPol surveys.  The modified gravity effects on the
mass function, halo bias, matter power spectrum, and mass-observable relation 
are taken into account.  We show that, relying on number counts only, the 
Planck cluster catalog is expected to reduce current upper limits by about 
a factor of four, to  $\sigma_{f_{R0}}=2\times10^{-5}$ (68\% confidence level) while SPT, SPTPol and ACTPol yield about $3\times10^{-5}$.  
Adding the
cluster power spectrum further improves the constraints to 
$\sigma_{f_{R0}}= 5 \times 10^{-6}$ for Planck and 
$\sigma_{f_{R0}}=2\times10^{-5}$ for SPTPol, pushing cluster constraints 
significantly beyond the limit where number counts have no constraining power 
due to the chameleon screening mechanism.  Further,
the combination of both observables breaks degeneracies, especially with the
expansion history (effective dark energy density and equation of state).  
The constraints are only mildly worsened by the use of 
self-calibration but depend on the mass threshold and redshift
coverage of the cluster samples.  
\end{abstract}
 
 
 
\maketitle

\section{Introduction} 
\label{sec:intro}

One of the most fascinating aspects of contemporary cosmology is the potential of 
constraining fundamental physics with the plethora of available data.  
Galaxy clusters constitute one of the major tools we can use to this aim.  
The biggest gravitationally bound objects in the Universe, they have
formed fairly recently and several of their global properties such as
abundance and clustering on large scales can be 
predicted accurately with theoretical models (e.g., \cite{Tin08,SheTor99}).  
For this reason, they have been extensively used in the past in order 
to constrain fundamental parameters such as the total matter density and 
the matter power spectrum normalization
\citep{Viana99,Pierpaoli01,Pierpaoli03,Borgani01,Allen03a}.
When combined with other cosmological data at various redshifts, clusters 
can also be used to constrain particle physics and neutrino properties \citep{Pierpaoli04,Allen03b,Vikhlinin09}.  

Given that gravity is the only relevant force in the formation of 
structure in the Universe on large scales, cosmological observations
are uniquely suited to test gravity on scales of Mpc, complementing
Solar System tests on AU scales.  
In recent years, clusters have received considerable interest as a probe
of gravity \cite{Martino09,Diaferio09}.  Modifications to gravity
generically change the growth of large-scale structure (e.g., \cite{JainKhoury,CliftonEtal}),
and clusters
at the high-mass tail of the mass function are especially sensitive
to changes in the growth rate.  This has been exploited in \citet{Schmidt:2009am} who
used a sample of X-ray clusters to constrain $f(R)$ gravity.  Similarly,
\citet{Lombriser:2010mp} used an optical Sloan cluster sample.  A consistency test
of the General Relativity + smooth Dark Energy framework 
using clusters was done in \cite{RapettiEtal}.  

Here, we focus on the $f(R)$ model of gravity, using the functional
form proposed in \cite{HuSaw07a}.  This model produces acceleration without
a true cosmological constant, and is indistinguishable from 
$\Lambda$CDM through geometric probes (CMB, Supernovae, $H_0$, BAO 
measurements).  However, gravitational forces are modified on smaller 
scales.  Furthermore, the model 
includes the chameleon screening mechanism which restores General Relativity
in high-density environments.  Thus, this model is able to
satisfy all current constraints on gravity.  Structure formation in this modified
gravity model is now understood on all cosmological scales: the linear regime 
of structure formation in this $f(R)$ model has been studied in 
\cite{HuSaw07a}.  The non-linear structure formation was investigated
using dedicated N-body simulations in \cite{oyaizu08b,Pkpaper,halopaper,LiEtal}.  
This allows for
fully self-consistent constraints and forecasts to be made for this
model.  

While cluster samples have mainly been selected in the optical 
and X--ray bands in the past, recent observations based on the Sunyaev--Zeldovich (SZ) 
effect are starting to produce new detections \cite{Williamson11,Vanderlinde10,Marriage11,Planck11a, Planck11b}.
The SZ effect consists in CMB photons inverse--Compton scattering off
electrons in the intra--cluster medium. This process causes a
distortion in the CMB blackbody spectrum, and a frequency-dependent
brightness change \cite{Birkinshaw99}.  
What makes SZ clusters particularly interesting as cosmological probes is 
the unique, almost redshift-independent sensitivity for detecting clusters.
As a consequence, SZ surveys have the potential to discover clusters at high redshift where
optical and X-ray surveys are not very efficient.
This new probe is  receiving significant attention because of additional data expected  from ongoing
SZ surveys like Planck, the Atacama Cosmology Telescope
(ACT), and the South Pole Telescope (SPT) in the near future.

In this paper, we explore to what extent these new cluster surveys are expected 
to constrain $f(R)$ models through cluster number counts and clustering.  
The paper is organized as follows.  
We begin by presenting the surveys and expected cluster samples in 
\refsec{data}.  This is useful as the modified gravity effects discussed
throughout the paper depend sensitively on the characteristics of the
cluster samples.  In \refsec{th} we present the
parametrization of modified gravity effects on the halo abundance 
and clustering.  \refsec{Fisher} details the Fisher formalism employed here,
as well as the fiducial cosmology adopted.    
The forecasted constraints are presented in \refsec{res}.  We discuss
our results in \refsec{disc} and conclude in \refsec{conc}.

\section{Cluster surveys}
\label{sec:data}

We will investigate the predictions for the four surveys described in the
following.  While we try to obtain as realistic survey specifications as
possible, in particular for the mass limit as function of redshift 
$M_{\rm lim}(z)$, the lack of previous large samples of SZ clusters
necessarily makes these quantities somewhat uncertain.  In particular,
the relation between cluster mass and SZ signal is still imperfectly known
(e.g.~\cite{Ameglio2009,Rasia2005,Nagai2007,PifVal2008}).  
 The final mass limits as a function of redshift are 
shown in \reffig{mlimz}, and the resulting expected number of clusters
for each survey is shown in \reffig{Nz}.  

\subsection{The \planck\ Catalog} 
\planck\ is imaging the whole sky with an unprecedented combination of sensitivity ($\Delta T/T\sim2\times10^{-6}$ per beam  at 100 - 217 GHz), angular resolution ($5'$ at 217 GHz), and frequency coverage ($30-857$ GHz). The SZ signal is expected to be detected from a few thousand individual galaxy clusters.  \planck\ will produce a cluster sample with median redshift $\sim0.3$ (see \reffig{Nz}, upper left panel). The SZ observable is the integrated Comptonization parameter $Y=\int y\ d\Omega_{\rm cluster}$  out to a given radius. For Planck,  a $5\sigma$  detection threshold ensuring high level of completeness (about 90\%)  corresponds to  $Y_{200,\rho_c}\ge2\times10^{-3}{\rm arcmin^2}$~\citep{Melin2006}, where $Y_{200,\rho_c}$ is 
the integrated comptonization parameter within $r_{200,\rho_c}$, the
radius enclosing a mean density of 200 times the critical density.  
The early release from the \planck\ Collaboration gives a sample of 189 high signal-to-noise SZ clusters with $\ge6\sigma$ detection. 
 It is therefore likely that our assumed detection threshold  will be 
 eventually reached in future data releases.
  For an SZ survey, its flux limit can be translated into a limiting mass 
by using simulation-calibrated scaling relations \cite{Sehgal2007}:

\begin{equation}
\frac{M_{\rm lim,200\rho_c}(z)}{10^{15}M_\odot } =\left[\left (\frac{D_A(z)}{{\rm Mpc}/h_{70}}\right)^2 E(z)^{-2/3} \frac{Y_{\rm 200, \rho_c}}{2.5\times10^{-4}} \right] ^{0.533}.
\label{eqn:szflux}
\end{equation}

In order to mitigate the effect of overestimation of unresolved clusters at low redshift, we further restrict $M_{\rm lim,200\rho_c}$ to be at least $10^{14}M_{\odot}$ at all $z$. 
With all these criteria, the \planck\ survey is expected to detect $\sim1000$ clusters.
The mass threshold we find with this approach is consistent with the one in \cite{Schafer2007}.
While we keep $Y_{\rm 200, \rho_c}=2\times10^{-3}{\rm arcmin^2}$ as our reference minimum value for presentation of the  main results, we will 
also discuss predictions for a lower mass threshold, corresponding to $Y_{\rm 200, \rho_c}=10^{-3}{\rm arcmin^2}$. With such threshold, the completeness of the  $S/N > 5$ sample is reduced to about 70\%
and the total number of clusters is 2700.

\subsection{SPT and SPTpol}
The SPT survey is currently observing the sky with a sensitivity of $18 \mu$K/arcmin$^2$ at 148 GHz, 218 GHz, and 277 GHz. This survey covers $\Omega\approx2500$ square degrees of the southern sky (between $20h\geq{\rm RA}\geq7h$, $-65^{\circ}\le\delta\le-30^{\circ}$) with a projected survey size and cluster mass limit well matched to the Stage III survey specification of the Dark Energy Task Force~\cite{Vanderlinde2010}. 
For the mass limits, we employ the calibrated selection function of the survey by~\cite{Vanderlinde2010}.  This is based on simulations and used to provide a realistic measure of the SPT detection significance and mass. Disregarding the scatter in the fitting parameters for this relation, we use here:
\be
\frac{M_{\rm lim,200\bar{\rho}}(z)}{5\times10^{14}M_{\odot}h^{-1}}=\left [ \left ( \frac{\sqrt{\xi^2-3}}{6.01} \right )\left ( \frac{1+z}{1.6} \right )^{-1.6} \right ]^{1/1.31}
\label{eqn:spt}
\ee

\noindent where $\xi$ is the detection significance. For the SPT survey, we take clusters detected at $\xi>5$ which ensure a 90\% purity level.  Currently, the SPT team is setting a low redshift cut at $z_{\rm cut}=0.3$ in 
their released cluster sample, due to difficulties in reliably distinguishing  low-redshift clusters from CMB fluctuations 
in single frequency observations. Nevertheless, with upcoming multi-frequency observations, a lower cut $z_{\rm cut}=0.15$ will likely be attained. We therefore apply this cut in our work.  
With this, the SPT survey is expected to detect $\sim500$ clusters.

In addition to this, we also consider the upcoming SPT polarization survey (hereafter SPTpol) which will have an increased sensitivity of $4.5\mu$K/arcmin$^2$ at 150 GHz for a 3 year survey and sky coverage of 625 square degrees.  We scaled the mass limits by a factor of $3.01/5.95$ in \refeq{spt} to match with the expected mass limits of SPTpol clusters (Benson 2011, private communication).  We again use $z_{\rm cut}=0.15$, resulting in a total expected number of $\sim1000$ clusters.  
While these are the limits we use for our main results, we also discuss outcomes that consider a lower mass limit, corresponding to $\xi = 4.5$ (80\% purity). With this mass limit, SPT would find 800 clusters and SPTPol about 1400 clusters.

\subsection{ACTpol}
The Atacama Cosmology Telescope (ACT) has been observing a portion of the southern sky since 2008 consisting of two strips of the sky, each 4 degrees wide in declination and 360 degrees around in right ascension, one strip is centered at $\delta=-5^{\circ}$, and the other is centered at $\delta=-55^{\circ}$~\cite{Sehgal2007}. With a sensitivity of $\approx35\mu$K/arcmin$^2$, only about 100 clusters are expected to be detected. Instead, we turn to the newly developing dual-frequency (150 GHz and 220 GHz) polarization sensitive receiver (hereafter ACTpol~\cite{Niemack2010} and reference therein) to be deployed on ACT in 2013. One of the three ACTpol observing seasons will have a wide survey covering $4000{\rm deg}^2$ to a target sensitivity of $20\mu$K/arcmin$^2$ in temperature at 150 GHz. With the wide field, they aim to find $\sim600$ clusters in the ACTpol survey.
The survey is $90\%$ complete above a limiting mass of $M_{\rm lim,200\bar{\rho}}=5\times10^{14}M_\odot h^{-1}$ (Sehgal 2011, private communication), and we therefore assume this as our redshift-independent mass limit for ACTpol.  As in SPT, the ACT team also put a low redshift cut in their parameter determination works and we likewise take $z_{\rm cut}=0.15$ for ACTpol, resulting in a total expected number of $\sim500$ clusters.  
We also present in the discussion section the results corresponding to a lower mass limit, 
$M_{\rm lim,200\bar{\rho}}=4\times10^{14}M_\odot h^{-1}$, which would result in a catalog of about 1000 clusters.

\begin{figure*}
  \begin{center}
       \includegraphics[width=85mm]{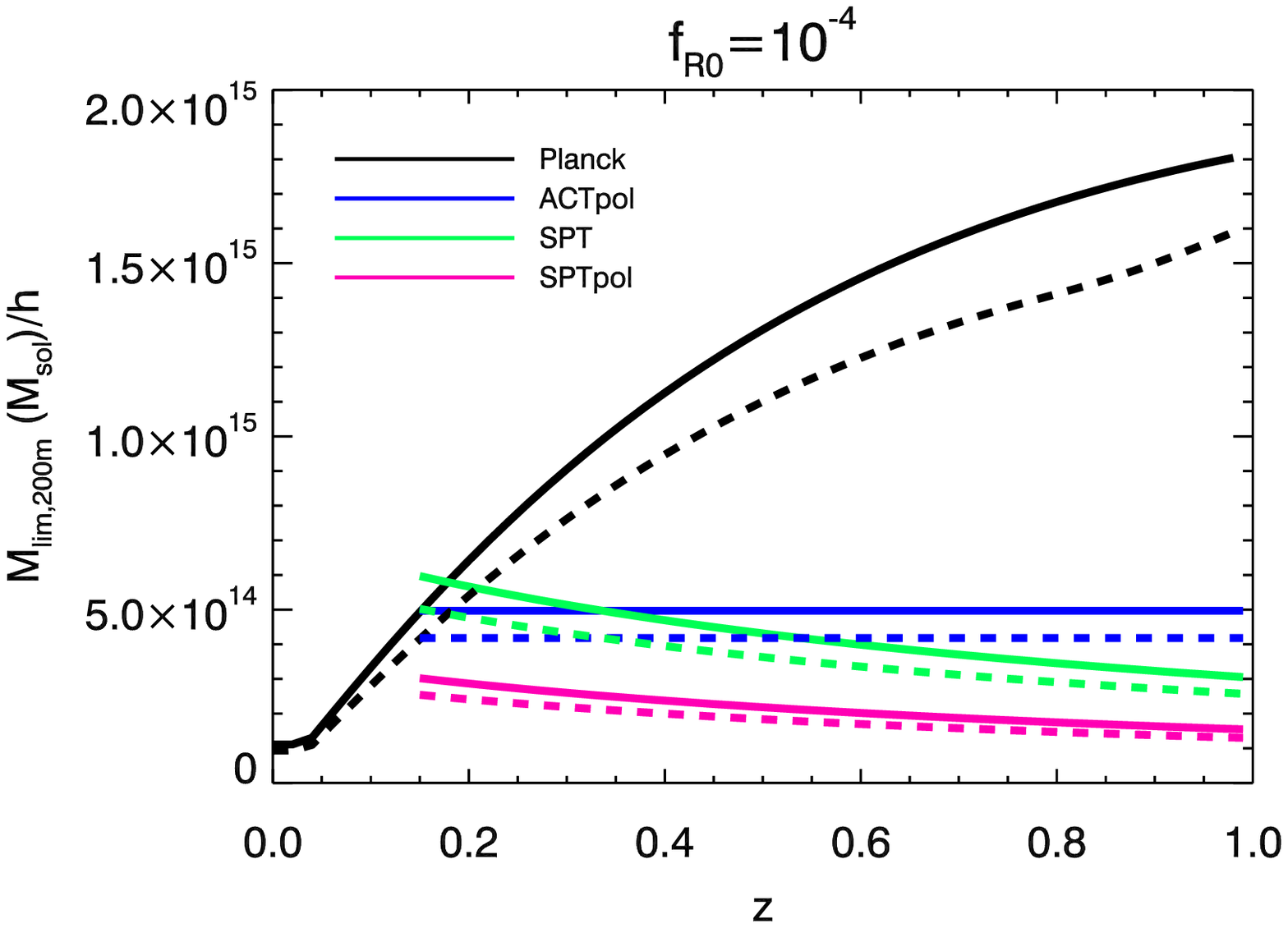} 
       \includegraphics[width=85mm]{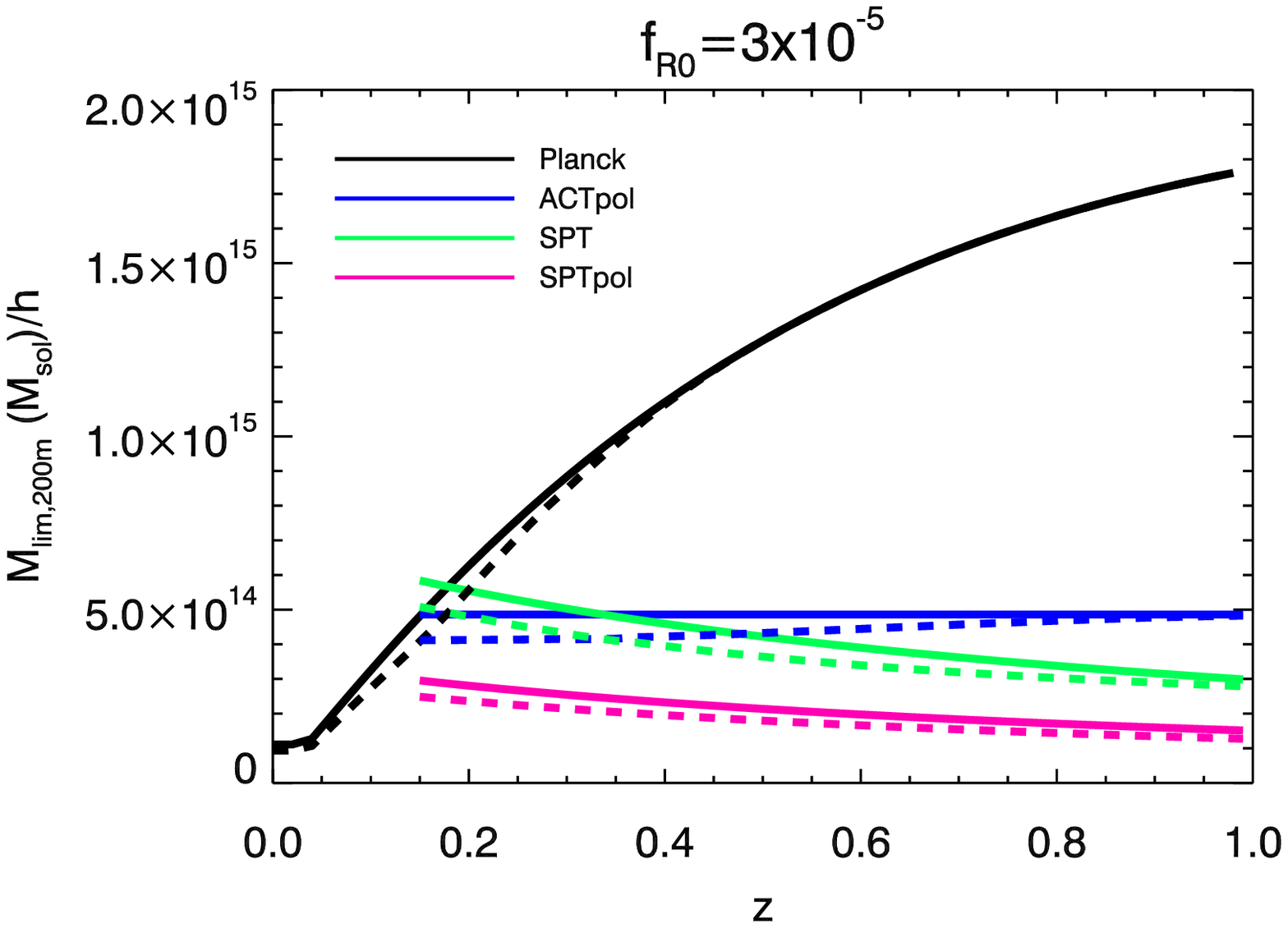}       
       \caption{Mass limit of cluster surveys in $\Lambda$CDM (solid) and $f(R)$ gravity (dashed) with $f_{R0}=10^{-4}$ (left) and $f_{R0}=3\times10^{-5}$ (right).  The mass limits in $f(R)$ are reduced due to the effect on dynamical
mass measurements (\refsec{fR}).
}
     \label{fig:mlimz}
  \end{center}
\end{figure*}

\section{Theoretical modeling}
\label{sec:th}

\subsection{$\bm{f(R)}$ gravity}
\label{sec:fR}

In the $f(R)$ model (see \cite{Nojiri:2006ri,Sotiriou:2008rp} and references therein), 
the Einstein-Hilbert action is augmented with a general function of the 
scalar curvature $R$ \cite{Caretal03,NojOdi03,Capozziello:2003tk},
 \begin{eqnarray}
S_{G}  =  \int{d^4 x \sqrt{-g} \left[ \frac{R+f(R)}{16\pi G}\right]}\,. 
\label{eqn:action}
\end{eqnarray}
Here and throughout $c=\hbar=1$.  This theory is equivalent to a 
scalar-tensor theory (if the function $f$ is nontrivial).  The additional
field given by $f_{R}\equiv df/dR$ mediates an attractive force whose
physical range is given by the Compton wavelength 
$\lambda_C= a^{-1}(3 d f_R/dR)^{1/2}$.  On scales smaller than $\lambda_C$,
gravitational forces are increased by 4/3, enhancing the growth of structure.   

A further important property of such models is the non-linear chameleon effect 
which shuts down the enhanced forces in regions with deep gravitational 
potential wells compared with the background field value,
$|\Psi| > | f_R(\bar R)|$ \cite{khoury04a,HuSaw07a}.  
This mechanism is necessary in order to pass
Solar System tests which rule out the presence of a scalar field locally.  
Thus, Solar System tests constrain the amplitude of the background field
to be less than typical cosmological potential wells today ($\sim 10^{-5}$).  

In this paper, we will choose the functional form introduced by Hu \& Sawicki
\cite{HuSaw07a}:
\begin{equation}
f(R) = - 2\Lambda \frac{R}{R+\mu^2},
\end{equation}
with two free parameters, $\Lambda$, $\mu^2$.  Note that as $R\rightarrow 0$,
$f(R)\rightarrow 0$, and hence this model does not contain a cosmological
constant.  
Nevertheless, as $R \gg \mu^2$,  the function $f(R)$ can be approximated as
\begin{equation}
f(R) = -2 \Lambda - f_{R0} \frac{\bar R_0}{ R} \,,
\label{eqn:fRapprox}
\end{equation}
with $f_{R0}= -2 \Lambda \mu^2/\bar R_0^2$ replacing $\mu$ as the second 
parameter of the model.  Here we define $\bar R_{0}=\bar R(z=0)$, 
so that $f_{R0}= f_{R}(\bar R_{0})$, where overbars denote the quantities of 
the background spacetime.  Note that $f_{R0} < 0$ implies $f_R < 0$ always, as 
required for stable cosmological evolution.  If $|f_{R0}| \ll 1$, 
the curvature scales set by $\Lambda ={\cal O}(R_0)$
and $\mu^2$ differ widely and hence the $R \gg \mu^2$ approximation is valid 
today and for all times in the past.

The background expansion history thus mimics $\Lambda$CDM with $\Lambda$ as
a true cosmological constant to order $f_{R0}$.   Therefore in the limit 
$|f_{R0}| \ll 10^{-2}$, the $f(R)$ model and $\Lambda$CDM are essentially 
indistinguishable with geometric tests.  The linear growth rate is
identical to that of $\Lambda$CDM on scales larger than $\lambda_C$, and 
becomes strongly scale-dependent on smaller scales \cite{HuSaw07a}.  

Note that we have chosen a model whose expansion history is close to 
$\Lambda$CDM by construction.  In general, there is sufficient freedom
in the free function $f$ to emulate any given expansion history
\cite{SonHuSaw07}.  Hence, below we will also allow the expansion history
to vary, parametrized by effective dark energy parameters $w_0$ and $w_a$.  
Further, while we choose a specific functional form for $f(R)$ here, it
is straightforward to map constraints onto different functional forms
(see \cite{FerraroEtal} for details).  In the following, for notational simplicity
$f_{R0}$ will always refer to the absolute value of the field amplitude today.  

\subsection{Cluster abundance in $\bm{f(R)}$}

Studying structure formation in $f(R)$ gravity beyond linear theory 
is complicated by
the non-linear field equation for the scalar field $f_R$, the non-linearity
being responsible for the chameleon mechanism.  The field equation needs
to be solved simultaneously with the evolution of the matter
density.  This has been done in the self-consistent N-body simulations
of \cite{oyaizu08b}.  The abundance of dark matter halos (mass function) and
their clustering (halo bias) in the $f(R)$ simulations was studied
in \cite{halopaper}.  

Since these simulations are very time-consuming,
they cannot be used to exhaust the cosmological parameter space.  Instead,
we use a simple model developed in \cite{halopaper} based on spherical 
collapse and the peak-background split in order to predict the 
cluster abundance and their linear bias.

In order to describe the effect of $f(R)$ gravity on the halo
mass function, we employ the Sheth-Tormen prescription
for the comoving number density of halos per logarithmic interval in the 
\emph{virial} mass $M_{\rm v}$, given by
\begin{align}
n_{\rm v}^{\rm (ST)} \equiv 
\frac{d n}{d\ln M_{\rm v}} &= {\bar \rho_{\rm m} \over M_{\rm v}} f(\nu) {d\nu \over d\ln M_{\rm v}}\,, 
         \label{eqn:massfn}
\end{align}
where the peak threshold $\nu = \delta_c/\sigma(M_{\rm v})$ and 
\begin{eqnarray}
\nu f(\nu) = A\sqrt{{2 \over \pi} a\nu^2 } [1+(a\nu^2)^{-p}] \exp[-a\nu^2/2]\,.
\end{eqnarray}
Here $\sigma(M)$ is the variance of the linear density field 
convolved with a top hat of radius $r$
that encloses $M=4\pi r^3 \bar \rho_{\rm m}/3$ at the background density
\begin{eqnarray}
\sigma^2(r) = \int \frac{d^3k}{(2\pi)^3} |\tilde{W}(kr)|^2 P_L(k)\,,
\label{eqn:sigmaR}
\end{eqnarray}
where $P_L(k)$ is the linear power spectrum (either
in $\Lambda$CDM or in $f(R)$) and $\tilde W$ is the Fourier transform
of the top hat window.  The normalization constant $A$ is chosen 
such that $\int d\nu f(\nu)=1$. The parameter values of $p=0.3$, $a=0.75$, and
$\delta_c=1.673$ for the spherical collapse threshold have previously been shown to 
match simulations of $\Lambda$CDM at the $10-20\%$ level. 
The virial mass is defined as the mass enclosed at 
the virial radius $r_{\rm v}$, at which the average density is $\Delta_{\rm v}$
times the mean density.   We transform the virial mass to the 
desired overdensity criterion $\Delta = 500/\Omega_m$ assuming a 
Navarro-Frenk-White 
\cite{NavFreWhi97} density profile \cite{HuKravtsov}, and assuming the
mass-concentration relation of \cite{Buletal01} (note that the rescaling
depends very weakly on the assumed halo concentration for the values of
$\Delta$ used here).  We thus obtain the mass function of halos
in the ST prescription, $n^{\rm (ST)}$, from $n_{\rm v}^{\rm (ST)}$.  

The effects of $f(R)$ modified gravity enter in two ways in this prescription:  
first, we use the linear power spectrum for the $f(R)$ model in 
\refeq{sigmaR}.  Second, we assume modified spherical collapse parameters
which were obtained by rescaling the gravitational constant by 4/3 during
the collapse calculation as well as the corresponding linear growth extrapolation
to obtain $\d_c$.    
This corresponds to the case where the collapsing region is always
smaller than the Compton wavelength of the field.  \citet{halopaper}
showed that this case always underestimates the $f(R)$ effects on the
mass function and bias, and hence serves as conservative model.  
For our fiducial cosmology at $z=0$, we
obtain GR collapse parameters of $\d_c = 1.675$, $\D_{\rm v} = 363$, while
the modified parameters are given by $\d_c = 1.693$, $\D_{\rm v} = 292$.  
The Sheth-Tormen prescription itself does not provide a very accurate
prediction for the abundance of clusters in $\Lambda$CDM in the entire redshift range relevant for SZ surveys.  Since
more precise parametrizations are available, we only use the
ST prescription to predict the relative \emph{enhancement} of the
cluster abundance in $f(R)$.  Specifically, after rescaling to our 
adopted mass definition, we take the ratio of the two and multiply it by
the $\Lambda$CDM mass function from \citet{Tin08},
\be
n(M,z)=n^{\rm (T)}_{\Lambda\rm CDM}(M,z)\frac{n^{\rm (ST)}_{f(R)}(M,z)}{n^{\rm (ST)}_{\Lambda\rm CDM}(M,z)},
\label{eqn:nn}
\ee
where we use the parameters given in their Appendix B.  
Note that for small field values and at high masses,
the predicted $f(R)$ mass function in fact becomes smaller than that for
$\Lambda$CDM.  Since
this effect is not seen in the simulations, we conservatively 
set the mass function ratio to 1 whenever it is predicted to be less than 1.  

\begin{figure*}
  \begin{center}
        \includegraphics[width=85mm]{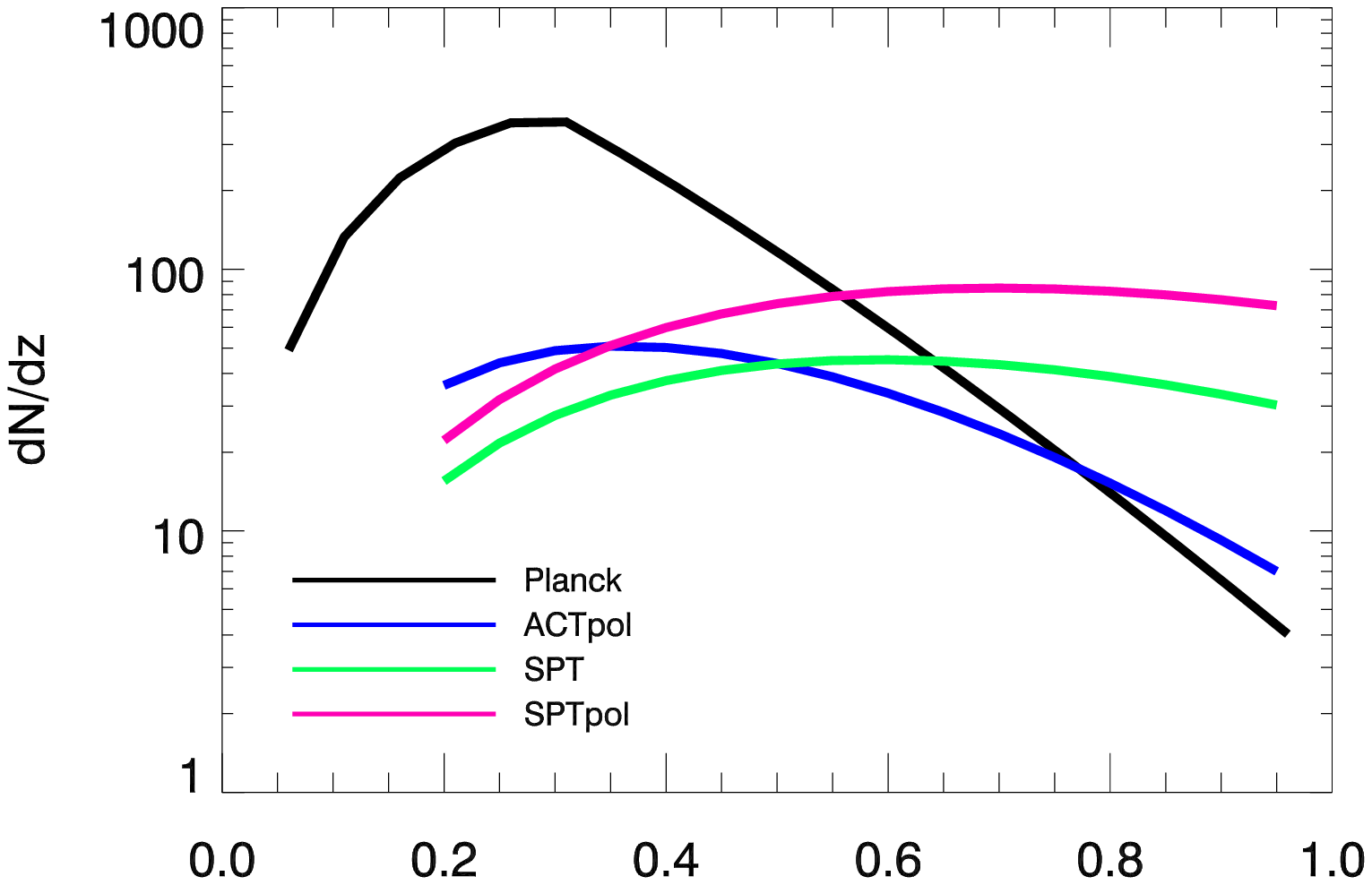} 
        \includegraphics[width=85mm]{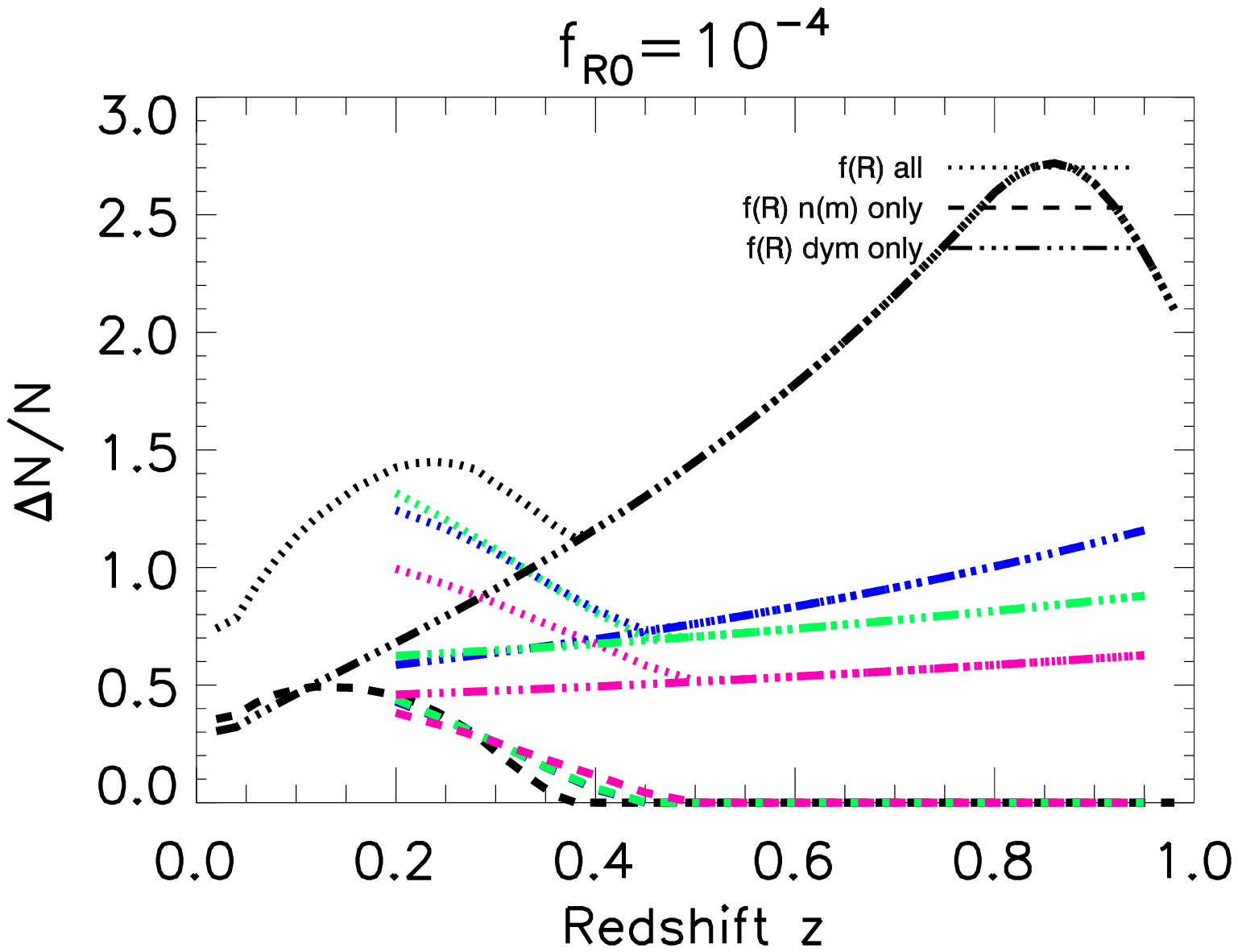} 
        \includegraphics[width=85mm]{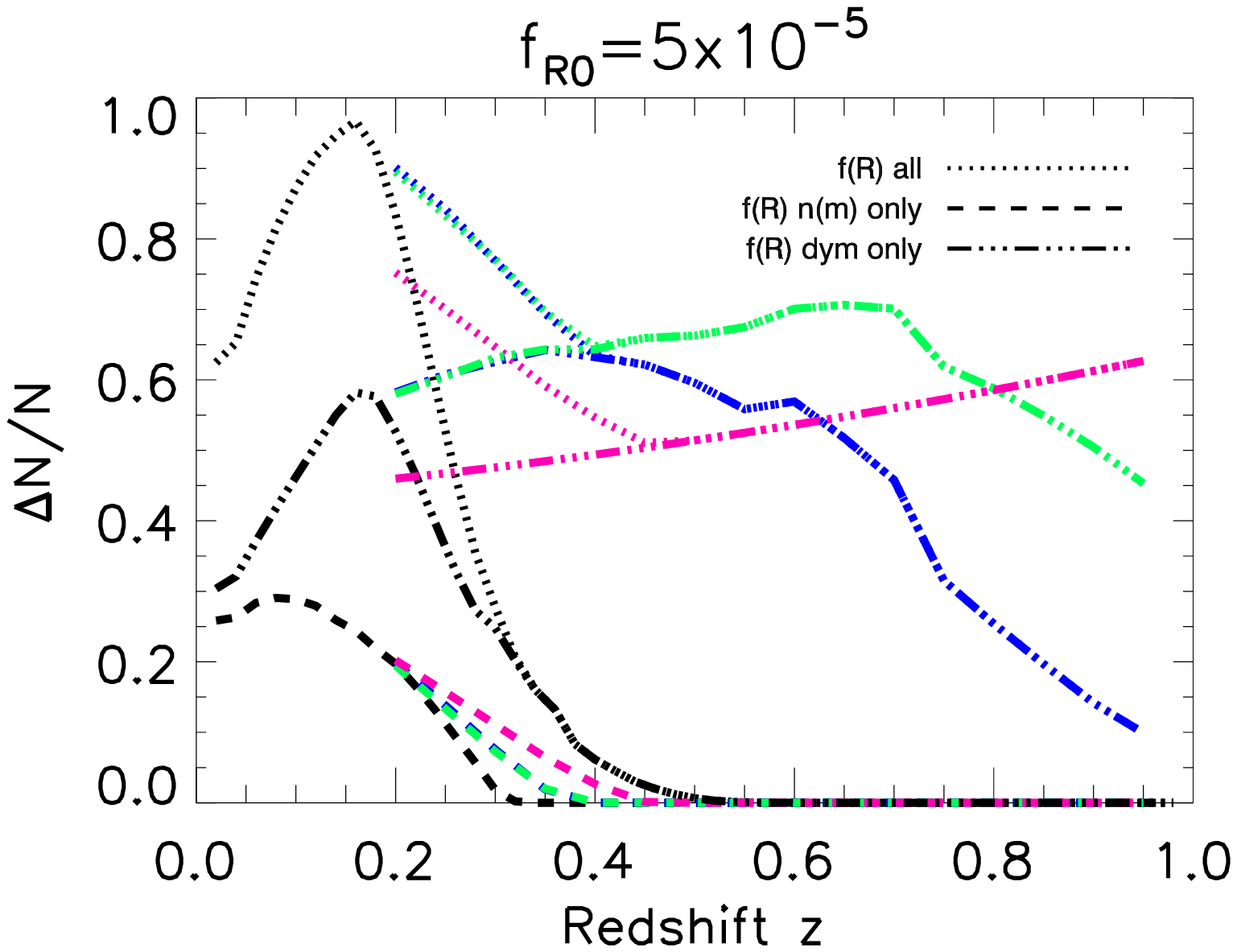}     
        \includegraphics[width=85mm]{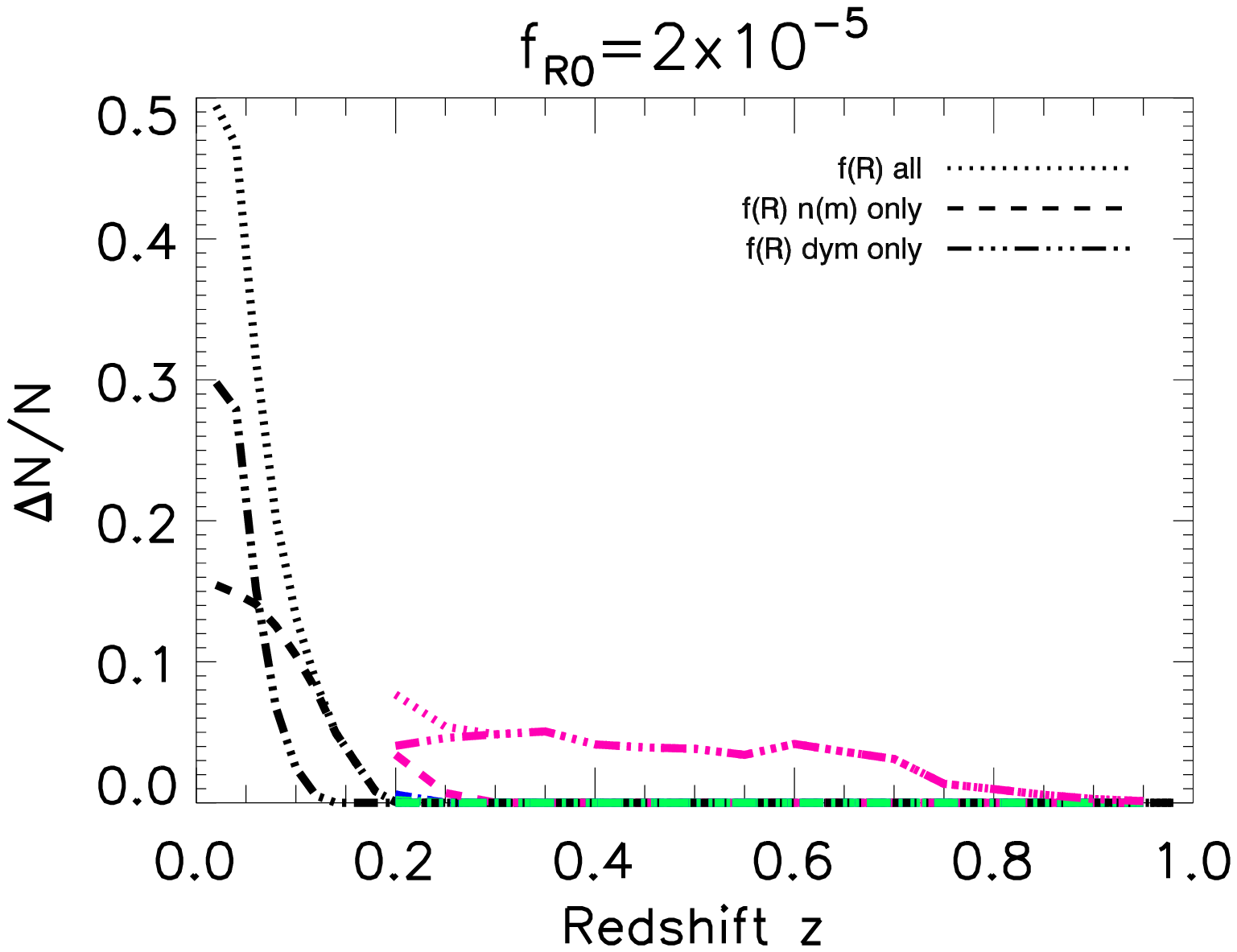}                   
       \caption{{\it Upper left.} The redshift distribution of clusters in the \planck\ (black), ACTpol (blue), SPT (green), and SPTpol (magenta) survey in the fiducial $\Lambda$CDM cosmology. {\it Upper right, bottom left, bottom right.} The fractional deviation of the number density between $f(R)$ and $\Lambda$CDM models due to all effects of $f(R)$ (dotted lines), $\Delta n_{\rm ln}$ effect only (dashed lines), and dynamical mass effect only (dot-dashed lines) evaluated at $f_{R0}=10^{-4}$, $f_{R0}=5\times10^{-5}$, and $f_{R0}=2\times10^{-5}$ respectively.  For the largest field value, the effect on the mass function dominates the enhancement of the cluster abundance at low $z$, while the dynamical mass effect dominates at $z \gtrsim 0.3$.  The \planck\ and SPTpol survey have the lowest mass threshold at $z<0.2$ and $z>0.2$ respectively and hence are most sensitive to the $f(R)$ effects for small field values.
}
     \label{fig:Nz}
  \end{center}
\end{figure*}

\reffig{Nz} shows the number of clusters as a function of redshift 
expected from the four surveys considered in this work (see \refsec{data}), 
and the relative deviations of the $f(R)$ model from the $\Lambda$CDM model
for different values of $f_{R0}$  (dashed lines).   
The $f(R)$ modifications
are most prominent at low redshifts $z \lesssim 0.4$, since the changes in the
linear power spectrum are restricted to progressively smaller scales towards
higher redshifts.  
Further, for $f_{R0} < 5\times 10^{-5}$, we see the strongest 
effects for surveys with the lowest mass thresholds, in particular 
Planck (for $z < 0.15$) and SPTpol.  
This is a consequence of the chameleon mechanism which suppresses the
mass function enhancement above progressively lower masses as $f_{R0}$
decreases.  There are negligible differences between $f(R)$ and 
$\Lambda$CDM for $f_{R0}<3\times10^{-5}$ at high halo masses.
  Hence, the mass threshold of a given survey determines what field values can be
probed by number counts.

Further, we have to take into account the effect of modified gravity
on the mass-observable relation.  The SZ effect is a dynamical mass
measure, as the decrement $Y$ is proportional to the velocity dispersion
(pressure) of electrons.  In modified gravity,
dynamical mass estimates are generally different from the actual
mass due to the presence of the additional gravitational force which
enters the virial equation.  As shown in \cite{dynamicalmass}, the
dynamical mass is related to the true mass via
\be
M_{\rm dyn} = \gbar^{3/5}\:M,
\label{eqn:Mdyn}
\ee
where $\gbar$ is a weighted integral of the force modification
over the object which describes the effect on the virial equation.  
In principle, $\gbar$ should be weighted by the
SZ emissivity and observational window function.  However in the interest of
simplicity, and since we are only interested in an approximate forecast,
we simply weight the modified forces by the matter density 
$\rho_{\rm NFW}(r)$ of the halo, assuming an NFW profile \cite{NavFreWhi97}.  
Further, we assume the host halo is spherically symmetric.  We then have
\be
\gbar = \frac{\int_0^{r_{\rm v}} dr\:r^2\:\rho_{\rm NFW}(r)\:\g(r)\:  r\, d\Psi_N/dr}
{\int_0^{r_{\rm v}} dr\:r^2\:\rho_{\rm NFW}(r)\: r\, d\Psi_N/dr},
\ee
where $\Psi_N$ is the Newtonian potential of the halo, found by
solving (see \cite{dynamicalmass} for an explicit expression)
\be
\nabla^2\Psi_N = 4\pi G\rho_{\rm NFW},
\ee
and $\g(r)$ is the force modification. In order to calculate the force
modification, we have to solve the chameleon field equation
for an NFW halo \cite{dynamicalmass}.  This calculation is computationally 
expensive, so we instead use a simple model which describes the exact results 
reasonably well \cite{dynamicalmass}; in fact it underpredicts the
exact result for the force modification, and thus is a conservative estimate.  
Specifically,
\be
\g(r) \approx 1 + \frac{1}{3}\frac{M(<r) - M(<r_{\rm scr})}{M(<r)}
\ee
Here, $r_{\rm scr}$ is the outermost radius at which the condition 
$|\Psi_N|\geq 3|\bar{f_R}|/2$ is met.  In the large-field limit this
condition is never met, so that $r_{\rm scr} = 0$ and $\g(r) = 4/3$
throughout.  \refeq{Mdyn} then yields $M_{\rm dyn}/M = (4/3)^{3/5} \approx 1.22$.  
For sufficiently small fields, the chameleon mechanism becomes active so that
$\g(r) \rightarrow 0$ for $r < r_{\rm scr}$, thus modeling the screening of 
the modified force.  In this case, $M_{\rm dyn}$ will interpolate between $M$ 
and $1.22 M$.  

We show in \reffig{mlimz} the mass threshold of the four cluster surveys 
in $\Lambda$CDM (solid) and the $f(R)$ dynamical mass effect to these 
thresholds (dashed).  \reffig{Nz} also shows the dynamical mass effect on the observed cluster abundance (dash-dotted lines).  
Note that the dynamical mass effect is not simply additive to the
mass function enhancement, since the latter depends on mass as well.  
Due to the steepness of the halo mass function at the high-mass end, the
fact that $M_{\rm SZ} = M_{\rm dyn}$ is larger than the true mass $M$ 
significantly boosts the abundance of detected clusters above the mass 
threshold.   The two effects of enhanced growth and increased $M_{\rm dyn}$ 
both contribute to increase the observed 
cluster abundance.  For $z \lesssim 0.4$, the mass function enhancement
provides a significant contribution to the overall change in number counts, 
while at higher redshifts the increase in dynamical mass is
the dominant effect.  

\subsection{Halo clustering in $\bm{f(R)}$}

In addition to the halo abundance, $f(R)$ modified gravity also affects 
the clustering of halos.  This effect comes from two sources:  first,
the matter power spectrum is enhanced on small scales by the increased
gravitational forces.  Second, the linear bias $b_L(M)$ of halos at a given mass
$M$ is reduced, since at a fixed mass halos are less rare in $f(R)$ than
in GR.  The power spectrum of clusters of mass $M$ is modeled as
\be
P_h(k,z|M) = b_L(M)^2 P_L(k,z).
\label{eqn:Ph}
\ee
The halo bias is given by the peak-background split.  For the mass function
used, it is given by \cite{SheTor99}
\be
b_L(M) = 1 + {a \nu^2 -1 \over \delta_c}
         + { 2 p \over \delta_c [ 1 + (a \nu^2)^p]}\,,
\label{eqn:bias}
\ee
where $\nu, a, p$ are defined after \refeq{massfn}.  Note that $\nu$ is given 
in terms of the virial mass $M_{\rm v}$ and thus, for a given
cluster mass $M$, it differs in the two collapse scenarios because of different $\D_{\rm v}$ values.  

\begin{figure*}
  \begin{center}          
       \includegraphics[width=85mm]{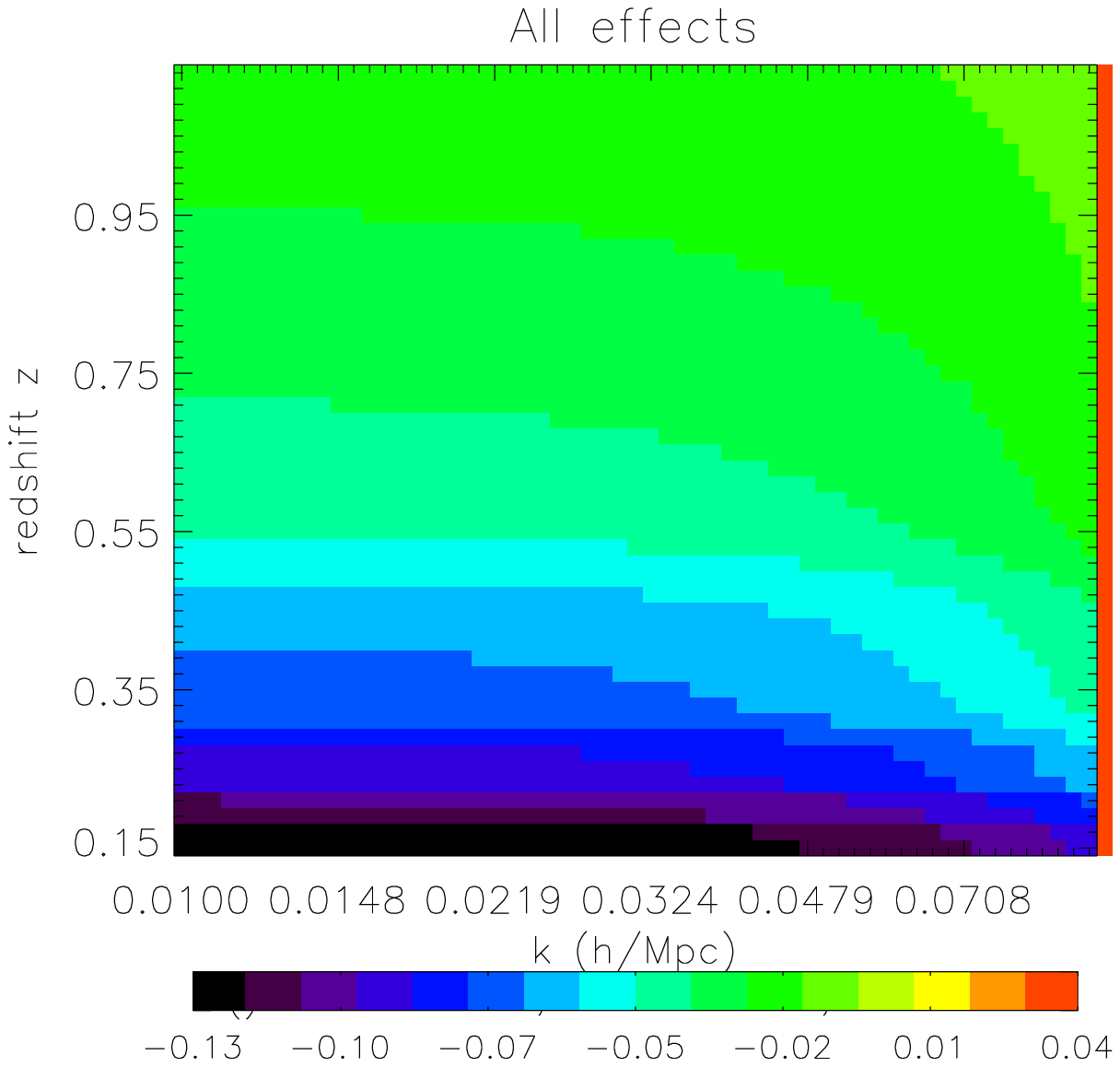} 
       \includegraphics[width=85mm]{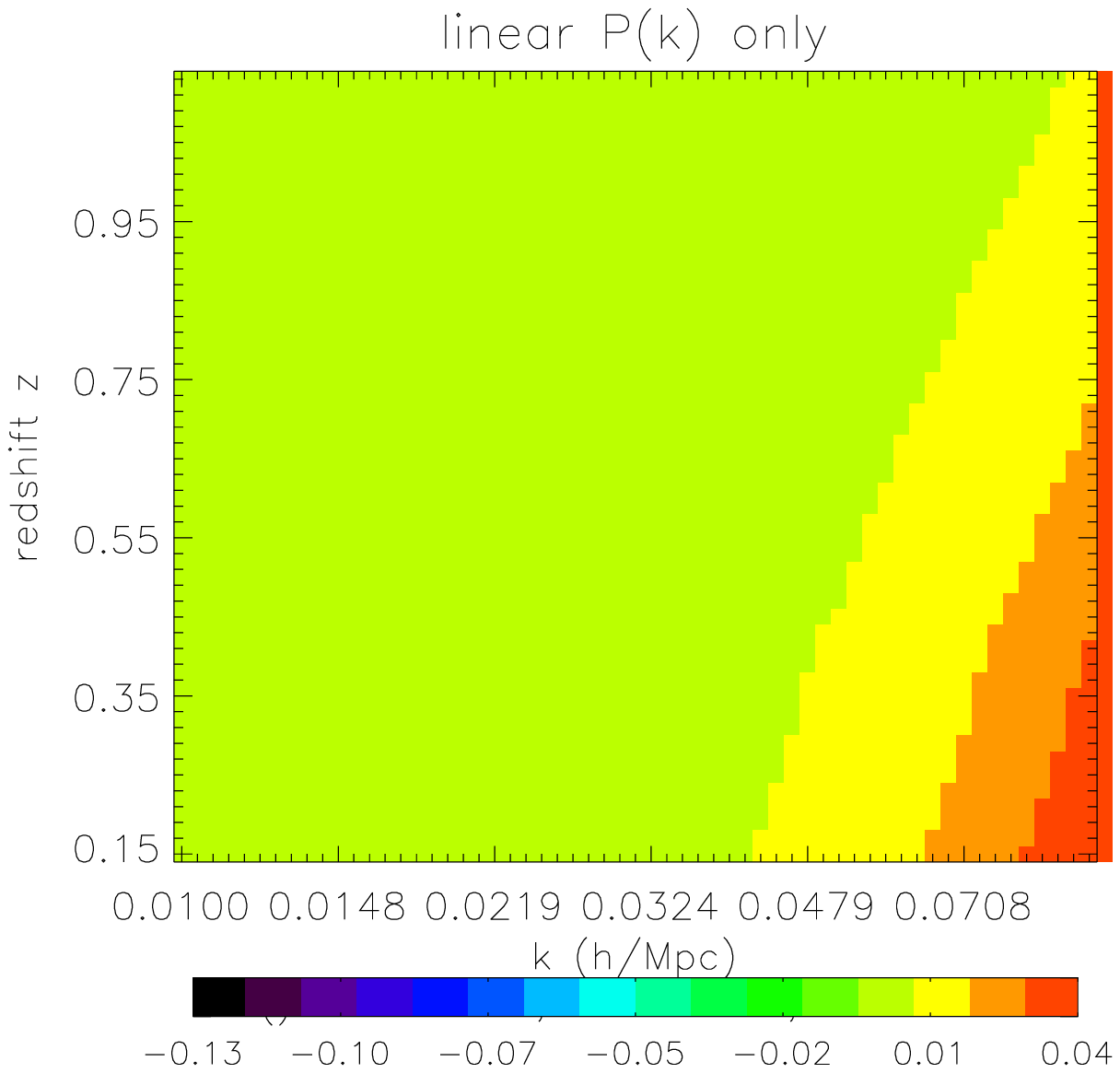}
       \includegraphics[width=85mm]{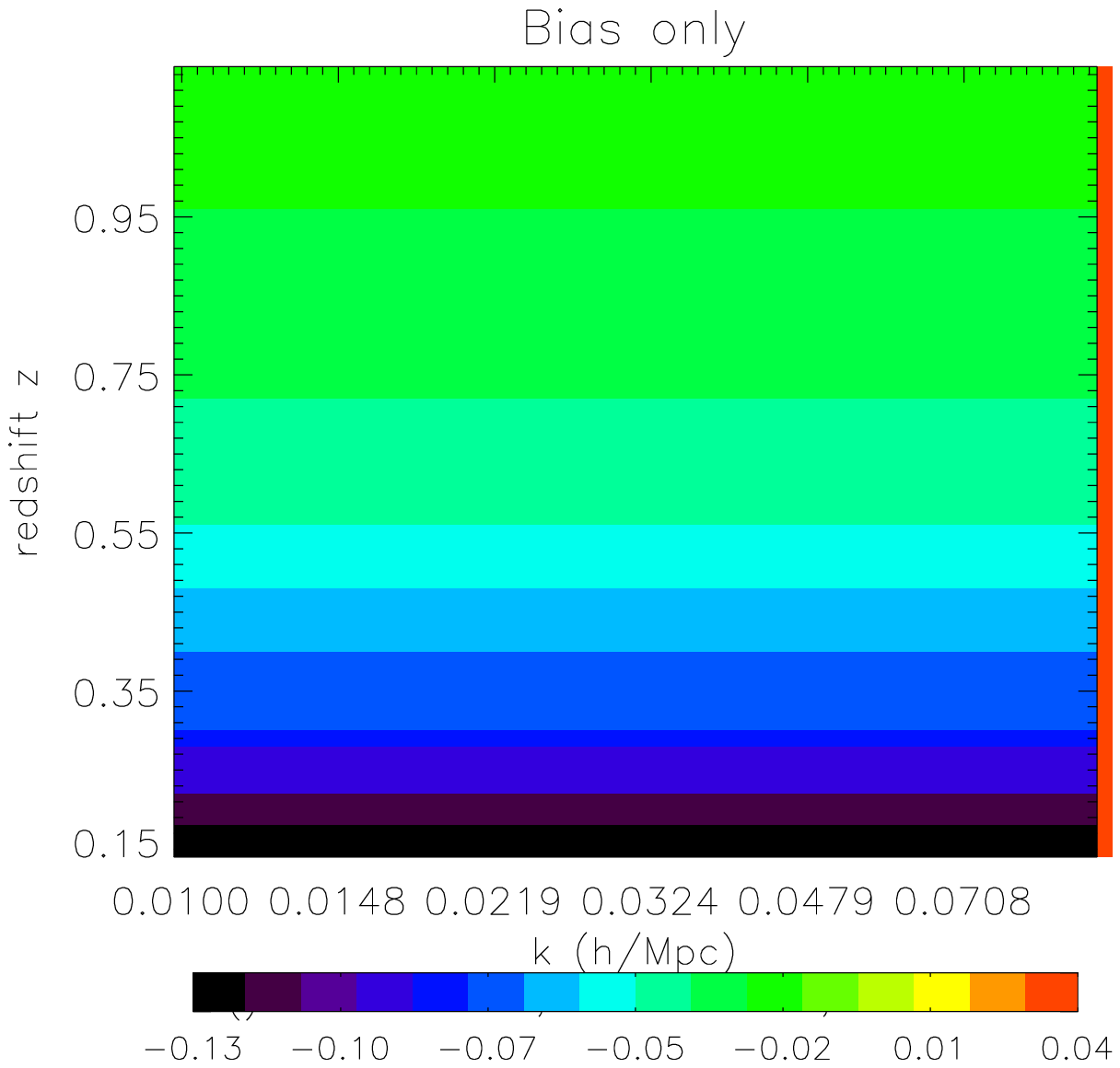}    
       \caption{
Relative deviations in the $f(R)$ halo power spectrum from $\Lambda$CDM, i.e. $\Delta P_h/P_h$ for the \planck\ survey, with $|f_{R0}|=10^{-5}$. {\it Upper left.} Total  deviation.
{\it Upper right.} Deviation due to $P_L(k)$ only. {\it Lower left.} Deviation due to halo bias $b_{L}$ only.  For this value of $f_{R0}$, the dynamical mass effect on the power spectrum is negligible and therefore we do not show it here. The redshift and scale dependence in the relative deviations from other cluster surveys are similar to the ones shown here.}
     \label{fig:deltap_planck1e-5}
  \end{center}
\end{figure*}

For the matter power spectrum in \refeq{Ph}, we use the linear theory
power spectrum for $f(R)$ and $\Lambda$CDM.  As shown in \cite{Pkpaper},
this describes the non-linear power spectrum at $z=0$ measured in $f(R)$ N-body
simulations up to scales $k\sim 0.2h/\rm Mpc$.  In order to minimize the
impact of non-linearities on the power spectrum and its covariance, we limit 
our Fisher matrix to modes with $k$ less than 0.1$h/$Mpc.  
Note that including
smaller scales will further improve the constraints; however, a more
sophisticated model including non-linear and/or scale-dependent bias, and
the non-linear matter power spectrum would be necessary in this case.  

Thus, the effect on the cluster power spectrum is due to three combined effects: enhancement of the linear power spectrum $\Delta P_{L}(k)$, halo bias $\Delta b_L(M,z)$, and the dynamical mass effect $M_{\rm dyn}$.  \reffig{deltap_planck1e-5} shows the relative deviation $\Delta P_h/P_h$ of the cluster power spectrum in $f(R)$ with respect to $\Lambda$CDM for the \planck\ survey (\refsec{data}) as a function of redshift and wavenumber $k$.  
Plots for the other surveys investigated here show similar $z-$ and 
$k-$dependences, though the amplitude of each effect depends on the survey.  
Here, we have assumed one mass bin $M > M_{\rm lim}(z)$ and $f_{R0} = 10^{-5}$.  
Similar to $dN/dz$, we plot the total effect (upper left), and separately the effect due to $\Delta P_L(k)$ (upper right), and $\Delta b_L(M,z)$ (lower panel). 
For this field value, the dynamical mass effect is irrelevant since the clusters
detectable by Planck are chameleon-screened. 
The departure from $\Lambda$CDM is mainly driven by $\Delta b_L(M,z)$ which shows a strong redshift dependence, and only mildly affected by $\Delta P_L(k)$ which is $k$-dependent and only relevant on small scales.  
Given that the power spectrum is shot-noise dominated at all scales for
the cluster samples considered, the effect on the linear halo bias in fact is the
most important contribution to the $f(R)$ constraints from the cluster
power spectrum.

\section{Fisher matrix formalism}
\label{sec:Fisher}

The Fisher information Matrix (FM hereafter) is defined as

\be
F_{\alpha\beta}\equiv -\left \langle \frac{\partial^2 \ln\mathfrak{L}}{\partial p_{\alpha} p_{\beta}} \right \rangle
\label{eqn:fisher}
\ee

\noindent where $\mathfrak{L}$ is the likelihood of a data set,
e.g. a cluster sample, written as a function of the parameters $p_\alpha$ describing
the model.  The parameters $p_\alpha$ comprise the cosmological model
parameters as well as ``nuisance'' parameters related to the data set
(e.g., mass calibration).  

\subsection{Cosmological parameters}

Throughout this paper, we assume a spatially flat ($\Omega_k=0$) cosmology.  
Our model comprises a total of seven cosmological parameters and one $f(R)$ model parameter which are left free to vary.  The seven parameters and their
fiducial values (in parenthesis, taken from the best-fit flat $\Lambda$CDM model from  
WMAP 7yr data, BAO and $H_0$ measurements \cite{KomatsuEtal11}) are: 
baryon density parameter $\Omega_bh^2$(0.0245); matter density parameter
$\omega_m \equiv \Omega_m h^2$ (0.143);
dark energy density $\Omega_\Lambda=1-\Omega_m$ (0.73);  power spectrum 
normalization $\sigma_8$ (0.809); index of power spectrum $n_s$ (0.963); 
effective dark energy equation of state through 
$w(z)=w_0+(1-a)w_a$, with fiducial values $w_0=-1$ and $w_a=0$.  
The Hubble parameter is then a derived parameter given by
$h = \sqrt{\omega_m/(1-\Omega_\Lambda)} = 0.73$ in the fiducial case.  
The $f(R)$ modification
can alternately be parametrized using the field amplitude $f_{R0}$ at $z=0$,
or the Compton wavelength $\lambda_{C0}$ at $z=0$ (see \refsec{LL}).  Our
fiducial value is $f_{R0} = \lambda_{C0} = 0$.  

In the following, we first discuss the Fisher matrix for number counts
and clustering of clusters, before describing the calibration parameters
and CMB priors.  Throughout, we divide the redshift range into bins $l$ of
width $\Delta z = 0.02$.  Further, we bin clusters in logarithmic mass bins $m$
of width $\Delta \ln M =0.3$  from
the minimum mass $M_{\rm lim}(z)$ for each survey 
(\refsec{data}) up to a large cut-off
mass of $M_{\rm max} = 10^{16} M_{\odot}$. 
Since the mass limit varies with redshift,
the number of mass bins thus also varies somewhat across the redshift
range.  

\subsection{Number counts}

The FM for the number of clusters $N_{l,m}$ within the $l$-th redshift bin and $m$-th mass bin is 

\be
F_{\alpha\beta}=\sum_{l,m}\frac{\partial N_{l,m}}{p_\alpha}\frac{\partial N_{l,m}}{p_\beta}\frac{1}{N_{l,m}}
\label{eqn:ncfisher}
\ee

\noindent where the sum over $l$ and $m$ runs over intervals in the whole redshift range $z=0-1$ and cluster mass range $[M_{\rm lim}(z),\infty]$. 
We can write the abundance of clusters expected in a survey, within a given redshift and mass interval, using the mass function as:

\bea
&& N_{l,m} = \Delta \Omega \Delta z  \frac{d^2V}{dzd\Omega} 
\int_{M_{l,m}}^{M_{l,m+1}} dM^{\rm ob} \\ \nonumber
&& \ \ \ \ \ \ \ \int_0^{\infty} d\ln M\: n(M,z) p(M^{\rm ob}| M)
\eea

\noindent where $\Delta\Omega$ is the solid angle covered by the cluster 
survey, $\ln M_{l,m} = \ln M_{\rm lim}(z_l) + m \Delta\ln M$, and 
$n(M, z)$ is the mass function given in \refeq{nn}.  
Following~\citet{LimHu05}, we take into account the intrinsic scatter in the relation between true and observed mass, as inferred from a given mass proxy, by the factor $p(M^{\rm ob}| M)$ which is the probability for a given cluster mass 
with $M$ of having an observed mass $M^{\rm ob}$. Under the assumption of a log--normal distribution for the intrinsic scatter, with variance $\sigma^2_{\ln M}$, the probability is 

\be
 p(M^{\rm ob}| M)=\frac{\exp [-x^2(M^{\rm ob})]}{\sqrt{2\pi\sigma^2_{\ln M}}}
 \ee
 
 \noindent where
 
 \be
 x(M^{\rm ob})=\frac{\ln M^{\rm ob}-B_{\rm M}-\ln M}{\sqrt{2\sigma^2_{\ln M}}}.
 \label{eqn:scal}
 \ee

\noindent With these notations, we parameterize the $M^{\rm ob}-M$ relation, in addition to the intrinsic scatter,  by a systematic fractional mass bias $B_{\rm M}$.  With this prescription, the final expression for the number count FM is:

\bea
N_{l,m} &=& \frac{\Delta \Omega \Delta z}{2}  \frac{d^2V}{dzd\Omega} \label{eqn:dndzfm}\\
& & \times \int_0^{\infty}  d\ln M\: n(M,z) 
\left ({\rm erfc}[x_m]-{\rm erfc}[x_{m+1}] \right),  \nonumber
\eea
where erfc$(x)$ is the complementary error function.  

\subsection{Power spectrum}

We define the FM for the power spectrum of galaxy clusters as
\begin{align}
F_{\alpha\beta}=\frac{1}{(2\pi)^2}\sum_{m,n} \sum_{l,i} &
\frac{\partial \ln P^{mn}_{\rm h}(k_i,z_l)}{\partial p_\alpha}
\frac{\partial \ln P^{mn}_{\rm h}(k_i,z_l)}{\partial p_\beta}\vs
& \times V_{l,i}^{mn,\rm eff} k_i^2 \Delta k
\label{eqn:psfisher}
\end{align}

\noindent where the sum over $m,n$ runs over mass bins, while the sum in $l$ and $i$ runs over intervals in the whole redshift range and wavenumber $0.01\ h{\rm Mpc^{-1}}\le k \le 0.1\ h {\rm Mpc^{-1}}$ with $\Delta \log_{10} k=0.017$ respectively. $P^{mn}_{\rm h}(k_i,z_l)$ is the cluster cross-power spectrum for mass bins $m$ and $n$, calculated for the given redshift and wavenumber through
\be
P^{mn}_{\rm h}(k_i, z_l) = b^m_{\rm eff}(z_l) b^n_{\rm eff}(z_l) P_L(k_i, z_l).
\ee
Here, $b^m_{\rm eff}$ is the mass function weighted effective bias, 
\be
b^m_{\rm eff}(z) = \frac{\int_{0}^{\infty} dM\:n(M,z)  b_L(M,z) ({\rm erfc}[x_m]- {\rm erfc}[x_{m+1}])}{\int_{0}^{\infty} dM\:n(M,z)  ({\rm erfc}[x_m]- {\rm erfc}[x_{m+1}])}.
 \label{eqn:beff}
\ee

\begin{figure}
  \begin{center}
       \includegraphics[width=80mm]{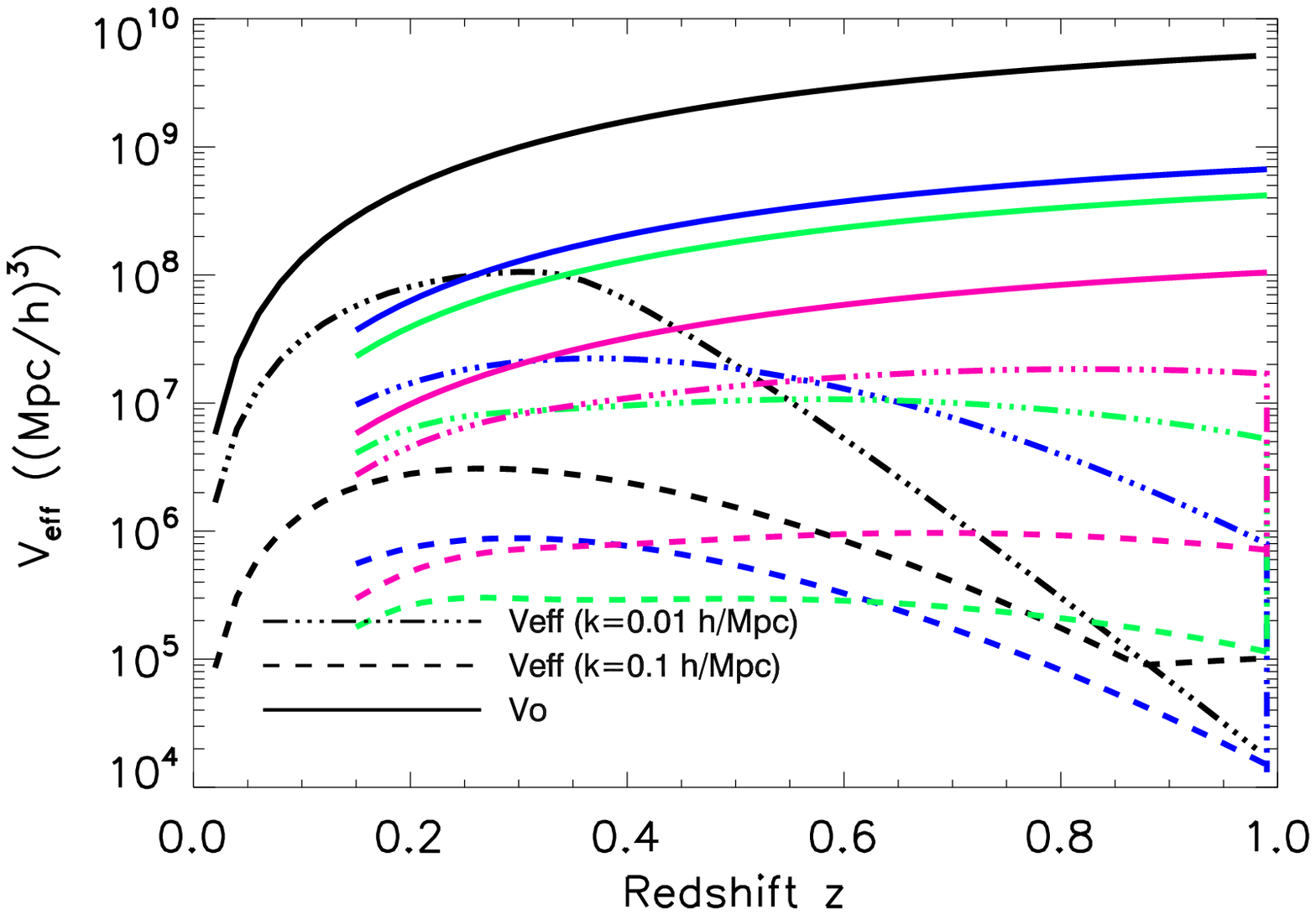} 
       \includegraphics[width=80mm]{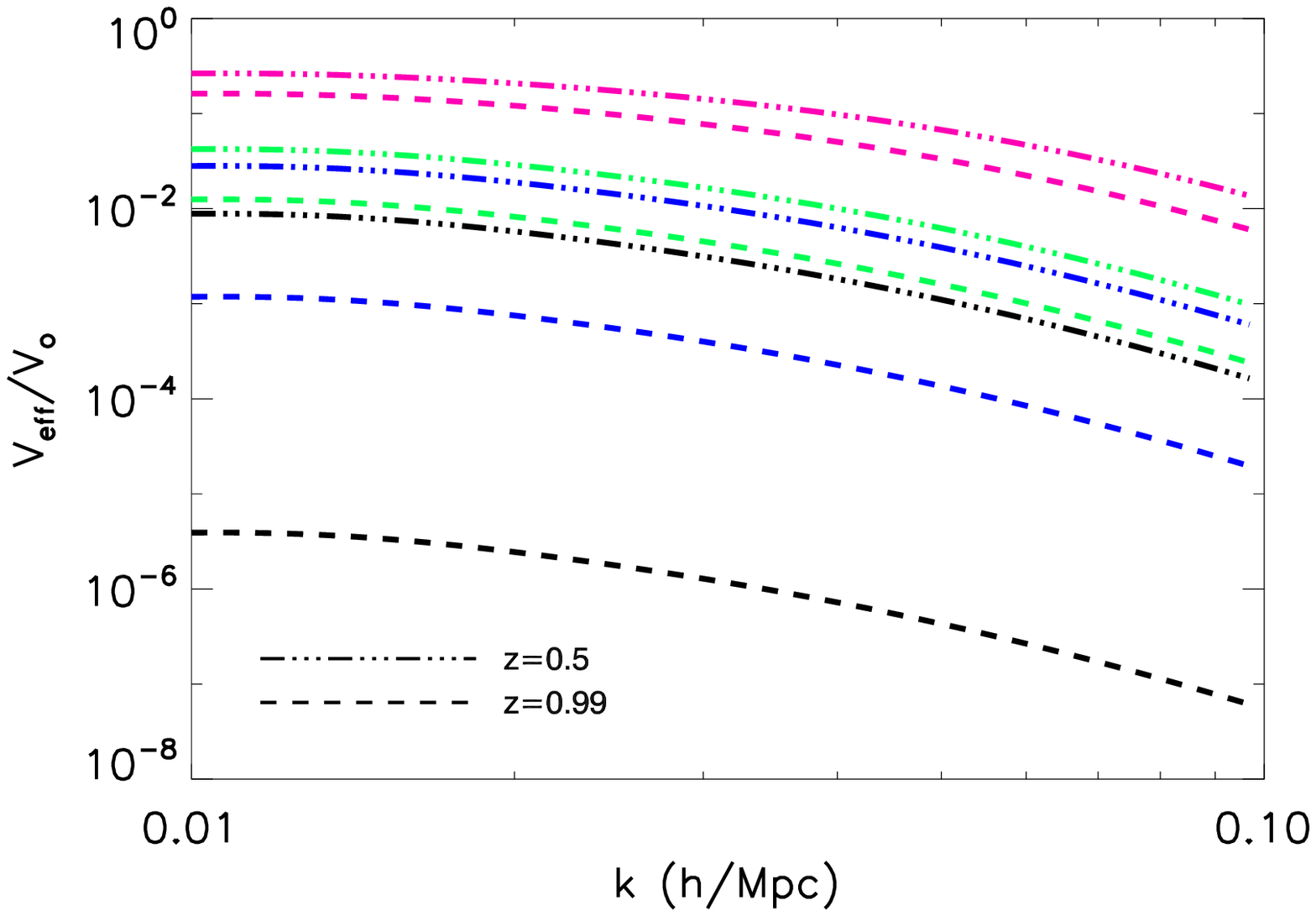}  
       \caption{The dependence on redshift {\it (top)} and wavenumber {\it (bottom)} of the effective volume (\refeq{veff}) for a single mass bin and each survey: \planck\ (black), SPT (green), SPTpol (magenta), and ACTpol (blue). The effective volume is a weak function of wavenumber $k$ but strongly depends on the redshift.}
     \label{fig:veff}
  \end{center}
\end{figure}

 \noindent The effective
volume for mass bins $m,n$, wave number $k_i$, and redshift $z_l$ is given by
(see App.~\ref{app:Pk})
 
\begin{align}
 \frac{V^{mn,\rm eff}(k_i, z_l)}{V_0(z_l)} =\:& [P^{mn}(k_i, z_l)]^2 n_m(z_l) n_n(z_l) \label{eqn:veff} \\
& \times \Big[(n_m P^{mm} + 1)(n_n P^{nn} + 1) \vs
&\quad\   + n_m n_n (P^{nm} + \delta^{nm} n_m^{-1})^2\Big]^{-1},\nonumber
\end{align}

 \noindent where $V_0(z)$ is the comoving volume of the redshift slice $[z_l-0.01,z_l+0.01]$ covered by the given survey, and $n_m(z_l)$ is the cluster number density for mass bin $m$ at redshift $z_l$. 
The effective volume gives the weight carried by each bin in the $(z,k)$ space to the power spectrum Fisher matrix, and hence quantifies the amount of information contained in a given redshift- and $k$-bin.  
\reffig{veff} shows the redshift and scale dependence of the effective volume for the four cluster surveys. 
We find that $V_{\rm eff} \lesssim 0.3\,V_0$ for all redshifts and surveys
considered, even when not binning in mass, hence the cluster power spectrum is shot-noise dominated for all surveys.   As the lower panel of \reffig{veff} illustrates, Planck is most limited by shot noise, while SPTpol is least limited, as expected from their respective mass limits and coverage.  

\subsection{Calibration parameters}

In self-calibrating the true and observed cluster mass (\refeq{scal}), we introduce four nuisance parameters which specify the magnitude and redshift-dependence of the fractional mass bias $B_M(z)$ and the intrinsic scatter $\sigma_{\ln M}(z)$.  Following~\cite{LimHu05}, we assume the following parametrization:

\bea
& B_M(z)=B_{M0}(1+z)^\alpha \nonumber \\
& \sigma_{\ln M}=\sigma_{\ln M,0}(1+z)^\beta
\eea
Therefore the four nuisance parameters are $B_{M0}$, $\alpha$, $\sigma_{\ln M,0}$, and $\beta$.  
A negative value for $B_M$ corresponds to an underestimation of mass.  The mass bias accounts for the possibility of a systematic offset in the calibration of the observable mass scaling relation.  
We adopt fiducial values of $B_{M0}=0$, $\alpha=0$,  $\sigma_{\ln M}=0.1$,  $\beta=0$.  
In deriving the main results, we will not make any assumption on the four nuisance parameters and leave them free to vary.  
We will study the effect of assuming different priors on the four nuisance parameters on the $f(R)$ constraints in \refsec{priors} . 

 \subsection{CMB Prior}
In the following, we present results with the Fisher matrix for the \planck\ CMB temperature power spectrum $C_l$ added to the constraints from cluster counts and power spectrum. We calculate the full CMB fisher matrix with CAMB~\cite{Lewis2000} and method described in~\cite{Pritchard2008}. For the \planck\ experiment, we use the three frequency bands 100, 143 and 217 GHz, and the $C_l$ are calculated up to $l_{\rm max}=2500$. Our fiducial parameter set for the CMB experiment is, as described in the DETF report~\cite{Albrecht2006},  $\theta=(n_s, \Omega_b h^2, \Omega_\Lambda, \Omega_m h^2, w_0, A_s, \tau)$, where $A_s$ is the primordial amplitude of scalar perturbations and $\tau$ is the optical depth due to reionization. After marginalizing over the optical depth, we transform the Planck CMB fisher matrix to our cluster survey parameter set ${\theta}'=(n_s, \Omega_b h^2, \Omega_\Lambda, \Omega_m h^2, w_0, \sigma_8)$ by using the appropriate Jacobian matrix. The CMB imposes strong prior on the cosmological parameters. For example, $\Omega_m h^2$ is known to be measured with the CMB power spectrum to an exquisite precision, and this helps in  breaking parameter degeneracies in the constraints from cluster surveys. As we shall see in \refsec{res}, the field amplitude parameter $f_{R0}$ shows degeneracies with some of the cosmological parameters, so that the CMB prior also helps in further constraining $f_{R0}$.

\subsection{Non-Gaussian likelihood}
\label{sec:LL}  

An inherent assumption in the Fisher matrix approach is that the
likelihood can be approximated as Gaussian around its maximum; 
in other words, that one can do a reasonably accurate Taylor 
expansion of $\ln\mathfrak{L}$ in all parameters.  Unfortunately, this is not
the case for the parameter $f_{R0}$, as the derivatives of the likelihood
with respect to $f_{R0}$ diverge at the fiducial value $f_{R0}=0$
(see Fig.~5 in \cite{Schmidt:2009am}).  Thus, we choose the Compton
wavelength $\lambda_{C0}$ as a parameter instead of $f_{R0}$, where
for the $f(R)$ model and fiducial cosmology considered here,
\be
\lambda_{C0}\approx32.53 \sqrt{\frac{|f_{R0}|}{10^{-4}}} \ {\rm Mpc}.
\label{eqn:lamc}
\ee
With this choice, $\ln\mathfrak{L}$ becomes analytic at the fiducial
value $\lambda_{C0}=0$.  Specifically,
we calculate the derivatives numerically as

\be
\frac{d\ln\mathfrak{L}}{d \lambda_{C0}}=\frac{\ln\mathfrak{L}(\lambda_{C0})-\ln\mathfrak{L}(0)}{\lambda_{C0}},
\ee
where $\lambda_{C0}$ is the Compton wavelength evaluated at the chosen 
step size $f_{R0} = \Delta f_{R0}$ through \refeq{lamc}, 
and $\mathfrak{L}$ denotes
the likelihood from either $dN/dz$ or $P(k)$.  
Unfortunately, the likelihood is still strongly
non-Gaussian in the direction of $\lambda_{C0}$, and the constraints
depend on the step size $\Delta f_{R0}$ chosen to evaluate
the Fisher matrix elements in \refeq{fisher}.  In principle, one would have
to evaluate the full likelihood with a MCMC approach, and then perform a  marginalization to
obtain proper forecasted constraints. 
Here, we opt instead for a simpler
approach.  
We evaluate the Fisher matrix for a range of step sizes $\Delta f_{R0}$, 
and then quote
the constraints for which $\sigma(f_{R0}) = \Delta f_{R0}$ is satisfied.  
One can easily show that this gives the correct answer in the ideal
case where the likelihood is Gaussian in all other parameters.  
Note that while we always use $\lambda_{C0}$ as parameter in the 
Fisher matrix,  we will  quote constraints in terms of $f_{R0}$ in order
to facilitate comparison with the literature, using  $\lambda_{C0}$ only to show parameter degeneracies.

\section{Results}
\label{sec:res}

We begin by discussing constraints from number counts (\refsec{NC}) and
power spectrum (\refsec{Pk}) separately, before moving on to
combined constraints (\refsec{comb}) and the impact of external priors
on the nuisance parameters (\refsec{priors}).  

\subsection{Number counts}
\label{sec:NC}

\begin{figure}
  \begin{center}
       \includegraphics[width=0.48\textwidth]{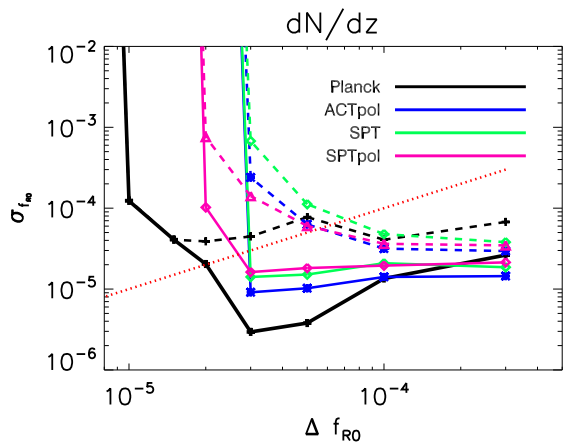}
       \caption{Fully marginalized 68\% confidence level (CL) constraints on $f_{R0}$ from the number count
of clusters only (using Planck CMB priors), as a function of the step size 
$\Delta f_{R0}$, for the surveys considered in this paper.  The red dotted
line indicates $\sigma_{f_{R0}} = \Delta f_{R0}$.  For a given survey, the 
intersection of this line with the predicted constraints yields the final
expected constraint (\refsec{LL}).  Solid (dashed) lines represent the case when dynamical mass is (is not) considered. The sharp upturn
at $\Delta f_{R0} \lesssim 5 \times 10^{-5}$ is due to the chameleon mechanism.
}
     \label{fig:NC}
  \end{center}
\end{figure}

As discussed in \refsec{LL},  the Fisher constraints depend on the value of
$\Delta f_{R0}$ adopted to evaluate the numerical derivatives in the Fisher
matrix.  \reffig{NC} shows the projected constraints for the
different surveys as a function of $\Delta f_{R0}$.  The sharp upturn
at $\Delta f_{R0} \sim 3 \times 10^{-5}$ (SPT and ACTPol),   $\Delta f_{R0} \sim 2 \times 10^{-5}$ (SPTPol) and $\Delta f_{R0} \sim 9\times 10^{-6}$ (Planck) signals the transition to the
chameleon-screened regime, where the mass function enhancement
becomes negligible \cite{halopaper}.   The shape of this 
transition depends on the mass limits of the different surveys, as more 
massive halos are screened for larger values of $f_{R0}$.  
The figure clearly shows that, with number counts alone,
constraints cannot be tighter than $\sigma_{f_{R0}} \sim 10^{-5}$.  Nevertheless,
this still constitutes an order of magnitude in improvement over current
constraints.  It should also be noted  that the use of the dynamical mass 
in the calculations leads to a significant improvement in constraints in the
large-field regime where the chameleon mechanism is not active.  

The precise constraints obtained at the intersection $\sigma_{f_{R0}} = \Delta f_{R0}$
are listed in \reftab{plancksigma}, along with the step size used
for each survey.  
The relative constraining power of the different surveys can easily be interpreted  by  looking at $\Delta N/N$ shown in \reffig{Nz}. The best  survey to constrain $f(R)$ with number counts is \planck\, 
which shows prominent deviations in $\Delta N/N$  
at low redshift, and yields a 68\% CL constraint of $\sigma_{f_{R0}}=2\times10^{-5}$.  
Although SPTpol shows significant differences in number counts out to large redshifts, the relatively small survey volume compared to Planck limits the performance in constraining $f(R)$ to $\sigma_{f_{R0}}\simeq  3 \times10^{-5}$.
It is interesting to notice that while their overall performance is similar, the constraints leverage on clusters in almost disjoint redshift ranges.  Therefore these surveys provide complementary information 
on $f(R)$ constraints from number counts, making the overall result less susceptible to specific issues related to either low or high redshift clusters.  
An investigation of whether the combination of both cluster samples 
yields a significant improvement on the expected $f(R)$ constraints would be worthwhile,
but is beyond the scope of this paper.
The other two surveys also  present  results highly competitive with current constraints, and not very different from SPTPol ($\sigma_{f_{R0}} = 3 \times10^{-5}$).
A better investigation with a proper likelihood would be necessary in order  to make more precise statements.

\begin{figure}
  \begin{center}
       \includegraphics[width=0.48\textwidth]{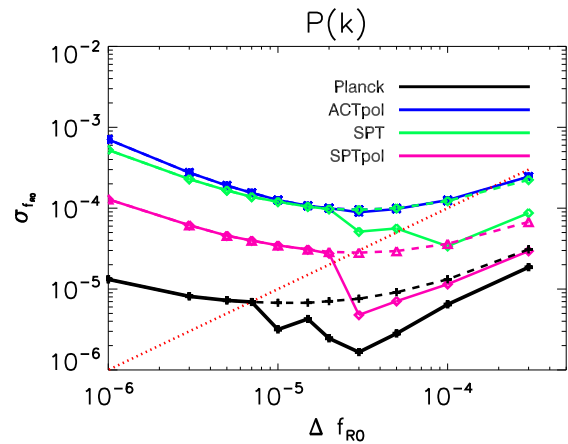}
       \caption{Same as \reffig{NC}, but from the power spectrum
of clusters only (using Planck CMB priors).  
}
     \label{fig:Pk}
  \end{center}
\end{figure}

\subsection{Power spectrum}
\label{sec:Pk}

\reffig{Pk} shows the constraints from the clustering of clusters alone
as a function of step size.  
The constraints generally worsen as the step size decreases to very small values. This is because the likelihood around the fiducial model ($\Lambda$CDM) scales as ${\lambda_{C0}}^a$, where $a>1$, and hence the derivatives go to zero as the step size decreases.  
However, constraints do not  worsen dramatically   as the step size crosses
the chameleon threshold, because the modification
to the halo bias in $f(R)$
persists even if the halos are chameleon screened \cite{halopaper}.
Furthermore, the
deviations in the matter power spectrum on small scales also persist
for field values $f_{R0} < 10^{-5}$.  As expected, the use of the dynamical mass
does not affect the  constraints for small field values
where the entire cluster sample is chameleon screened.  

The  constraints from power spectrum only are summarized in the second column of
\reftab{plancksigma}.  
 For Planck (as well as marginally for SPTPol) the
constraint on $f_{R0}$ from the cluster power spectrum is tighter
than that from the abundance only.  This is mainly because the
power spectrum retains sensitivity to $f(R)$ effects even when the
halos are chameleon screened.  
For ACTPol and SPT  the power spectrum yields slightly less constraining power than number counts, as the disadvantage of not having all-sky coverage  is not compensated by the relatively low mass threshold.

\begin{figure}
  \begin{center}
\includegraphics[width=0.48\textwidth]{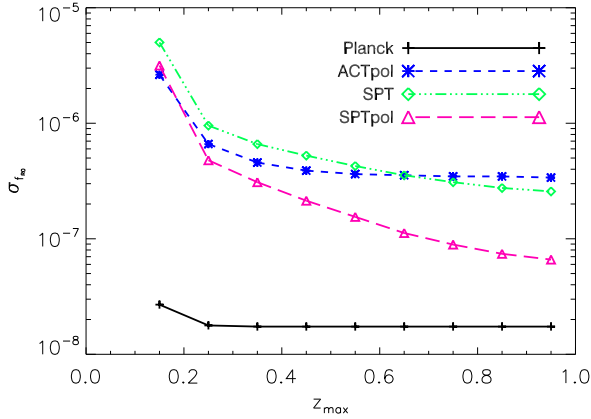}
       \caption{Fully marginalized constraints on $f_{R0}$ from the
power spectrum of clusters only, 
as a function of maximum cluster redshift $z_{\rm max}$.  
$\Delta f_{R0} = 10^{-5}$ was used for all values shown here.}
     \label{fig:sigz}
  \end{center}
\end{figure}

In order to investigate what cluster redshift range contributes to the
$f_{R0}$ constraints, we the constraints (for $\Delta f_{R0} = 10^{-5}$ fixed) 
as function of the maximum cluster redshift considered in \reffig{sigz}.  
For surveys with mass limits which decrease with redshift, i.e. SPT and 
SPTpol, constraints improve up to $z_{\rm max} = 1$, while for Planck all the 
information is derived from clusters below $z \approx 0.3$, 
and for ACT the constraining power comes from clusters below $z\approx 0.5$.  
It is especially interesting to compare results from ACTpol and SPT, which 
detect a comparable number of clusters overall but with a different redshift distribution.  \reffig{Nz} shows that ACTpol has a 
significantly higher number of clusters than SPT out to $z \approx 0.5$, and
a lower mass limit out to $z \approx 0.3$.  Yet the constraints from the
cluster power spectrum are worse for ACTpol than SPT, due to the contribution
from $z > 0.5$ clusters for SPT (\reffig{sigz}).  
How well each survey can realize their potential constraining power 
clearly depends on the precise $M_{\rm lim}(z)$ achieved in the final cluster sample.

\begin{table}
\caption{Marginalized constraints (68\% confidence level) from the two cluster probes  $dN/dz$ and $P(k)$, as well as the combination of both for the four SZ surveys.  The results are combined with forecasted constraints from the \planck\ CMB.  We also indicate the step size $\Delta f_{R0}$ used for each survey and probe.}
\begin{center}
\begin{tabular}{l ccc}
\hline\hline
Parameter & $dN/dz$  & $P(k)$  & $dN/dz$ + $P(k)$     \vspace*{0.05cm} \\ 
\hline\hline
& \multicolumn{3}{c}{Planck}\\
$\Delta f_{R0}$      & $2\times10^{-5}$     & $5\times10^{-6}$  & $5\times10^{-6}$ \\
\hline
       $f_{R0}$ &$    2.04\times10^{-5}$&$    7.24\times10^{-6}$&$    4.78\times10^{-6}$\\
    $\Omega_M h^2$ &$    1.10\times10^{-3}$&$    1.09\times10^{-3}$&$    1.07\times10^{-3}$\\
  $\Omega_\Lambda$ &$    0.17$&$    9.84\times10^{-2}$&$    3.18\times10^{-2}$\\
        $\sigma_8$ &$    6.31\times10^{-3}$&$    6.26\times10^{-3}$&$    6.15\times10^{-3}$\\
    $\Omega_b h^2$ &$    1.31\times10^{-4}$&$    1.30\times10^{-4}$&$    1.29\times10^{-4}$\\
             $n_s$ &$    3.33\times10^{-3}$&$    3.30\times10^{-3}$&$    3.27\times10^{-3}$\\
             $w_0$ &$    0.64$&$    0.37$&$    0.124$\\
             $w_a$ &$    2.34$&$    13.60$&$    1.12$\\
\hline
& \multicolumn{3}{c}{ACTpol}\\
$\Delta f_{R0}$     & $3\times10^{-5}$    & $10^{-4}$  & $2\times10^{-5}$ \\
\hline
          $f_{R0}$ &$    \sim3\times10^{-5}$&$    1.26\times10^{-4}$&$    \sim3\times10^{-5}$\\
    $\Omega_M h^2$ &$    1.10\times10^{-3}$&$    1.10\times10^{-3}$&$    1.09\times10^{-3}$\\
  $\Omega_\Lambda$ &$    0.17$&$    0.17$&$    0.13$\\
        $\sigma_8$ &$    6.31\times10^{-3}$&$    6.32\times10^{-3}$&$    6.27\times10^{-3}$\\
    $\Omega_b h^2$ &$    1.31\times10^{-4}$&$    1.31\times10^{-4}$&$    1.31\times10^{-4}$\\
             $n_s$ &$    3.33\times10^{-3}$&$    3.33\times10^{-3}$&$    3.30\times10^{-3}$\\
             $w_0$ &$    0.65$&$    0.64$&$    0.50$\\
             $w_a$ &$    2.18$&$    13.00$&$    1.52$\\
\hline
& \multicolumn{3}{c}{SPT}\\
$\Delta f_{R0}$      & $3\times10^{-5}$    & $5\times10^{-5}$  & $2\times10^{-5}$ \\
\hline
          $f_{R0}$ &$    \sim3\times10^{-5}$&$    5.62\times10^{-5}$&$    \sim3\times10^{-5}$\\
    $\Omega_M h^2$ &$    1.10\times10^{-3}$&$    1.09\times10^{-3}$&$    1.09\times10^{-3}$\\
  $\Omega_\Lambda$ &$    0.17$&$    8.82\times10^{-2}$&$    6.47\times10^{-2}$\\
        $\sigma_8$ &$    6.31\times10^{-3}$&$    6.28\times10^{-3}$&$    6.26\times10^{-3}$\\
    $\Omega_b h^2$ &$    1.31\times10^{-4}$&$    1.31\times10^{-4}$&$    1.31\times10^{-4}$\\
             $n_s$ &$    3.33\times10^{-3}$&$    3.30\times10^{-3}$&$    3.29\times10^{-3}$\\
             $w_0$ &$    0.65$&$    0.33$&$    0.24$\\
             $w_a$ &$    2.28$&$    5.56$&$    0.98$\\
\hline
& \multicolumn{3}{c}{SPTpol}\\
$\Delta f_{R0}$      & $2\times10^{-5}$     & $2\times10^{-5}$  & $2\times10^{-5}$ \\
\hline
          $f_{R0}$ &$    \sim3\times10^{-5}$&$    2.75\times10^{-5}$&$    1.74\times10^{-5}$\\
    $\Omega_M h^2$ &$    1.10\times10^{-3}$&$    1.09\times10^{-3}$&$    1.09\times10^{-3}$\\
  $\Omega_\Lambda$ &$    0.17$&$    6.06\times10^{-2}$&$    3.63\times10^{-2}$\\
        $\sigma_8$ &$    6.32\times10^{-3}$&$    6.26\times10^{-3}$&$    6.24\times10^{-3}$\\
    $\Omega_b h^2$ &$    1.31\times10^{-4}$&$    1.31\times10^{-4}$&$    1.30\times10^{-4}$\\
             $n_s$ &$    3.33\times10^{-3}$&$    3.29\times10^{-3}$&$    3.29\times10^{-3}$\\
             $w_0$ &$    0.62$&$    0.23$&$    0.14$\\
             $w_a$ &$    2.16$&$    2.96$&$    0.88$\\
\hline
\end{tabular} 
\end{center}
\label{t:plancksigma}
\end{table}

\subsection{Combined constraints}
\label{sec:comb}

\begin{figure}
  \begin{center}
       \includegraphics[width=0.48\textwidth]{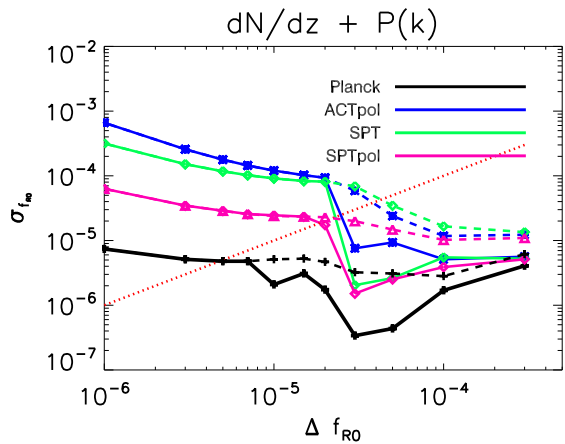} 
       \caption{Combined $dN/dz$ + $P(k)$ 68\% CL marginalized constraints on $f_{R0}$ as a function of the step size $\Delta f_{R0}$ for the different surveys.  As in \reffig{NC}, the red dotted line shows the identity $\sigma_{f_{R0}}=\Delta f_{R0}$.  
     \label{fig:comb}}
  \end{center}
\end{figure}

\reffig{comb} shows  constraints on $f_{R0}$ when combining both number
counts and clustering, as a function of the step size $\Delta f_{R0}$.
  The dependence on $\Delta f_{R0}$ is  similar
to the case of power spectrum-only and number counts-only constraints  at small and large  step size respectively.
Combining the two probes helps to break degeneracies and better
constrain the nuisance parameters.  As a result, the constraints on  $f_{R0}$ show improvements with respect to those derived from power spectrum or number counts  alone (third column in \reftab{plancksigma}).  
While  Planck reaches a constraint of $\sigma_{R0} \approx  5\times10^{-6}$ , ACTpol, SPT and SPTPol achieve $2-3 \times 10^{-5}$.
  Among the four surveys, the Planck survey thus yields the tightest constraints regardless of which cluster probe is being used.  
The relative merit of the Planck survey is due to its large area, which allows to detect massive clusters 
on the whole sky,  and its ability to detect  low redshift clusters.  
\reffig{Nz} shows that in the small-field regime ($f_{R0} \sim 10^{-5}$), the low redshift clusters drive the constraints for Planck, while low mass clusters do so for SPTPol.

Up to now, we presented results with conservative mass limits, i.e. clusters are expected to be detected with $S/N \ge 5$ for all the surveys. We also examined improvements in the constraint   $\sigma_{f_{R0}}$ when using more optimistic mass limits for each survey, according to what is outlined in section \refsec{data}.  
For all surveys, the constraints from number counts only are hardly affected,
since they are mainly set by the chameleon threshold.  In each case,
the larger cluster sample does improve the power spectrum constraints.  
However, only for Planck does this yield a significant improvement in 
the combined constraints (by a factor of 1.5 to $3\times 10^{-6}$), while
for ACTpol, SPT, and SPTpol, the improvement in combined constraints is 
marginal.

\begin{figure*}
  \begin{center}
\includegraphics[width=0.48\textwidth]{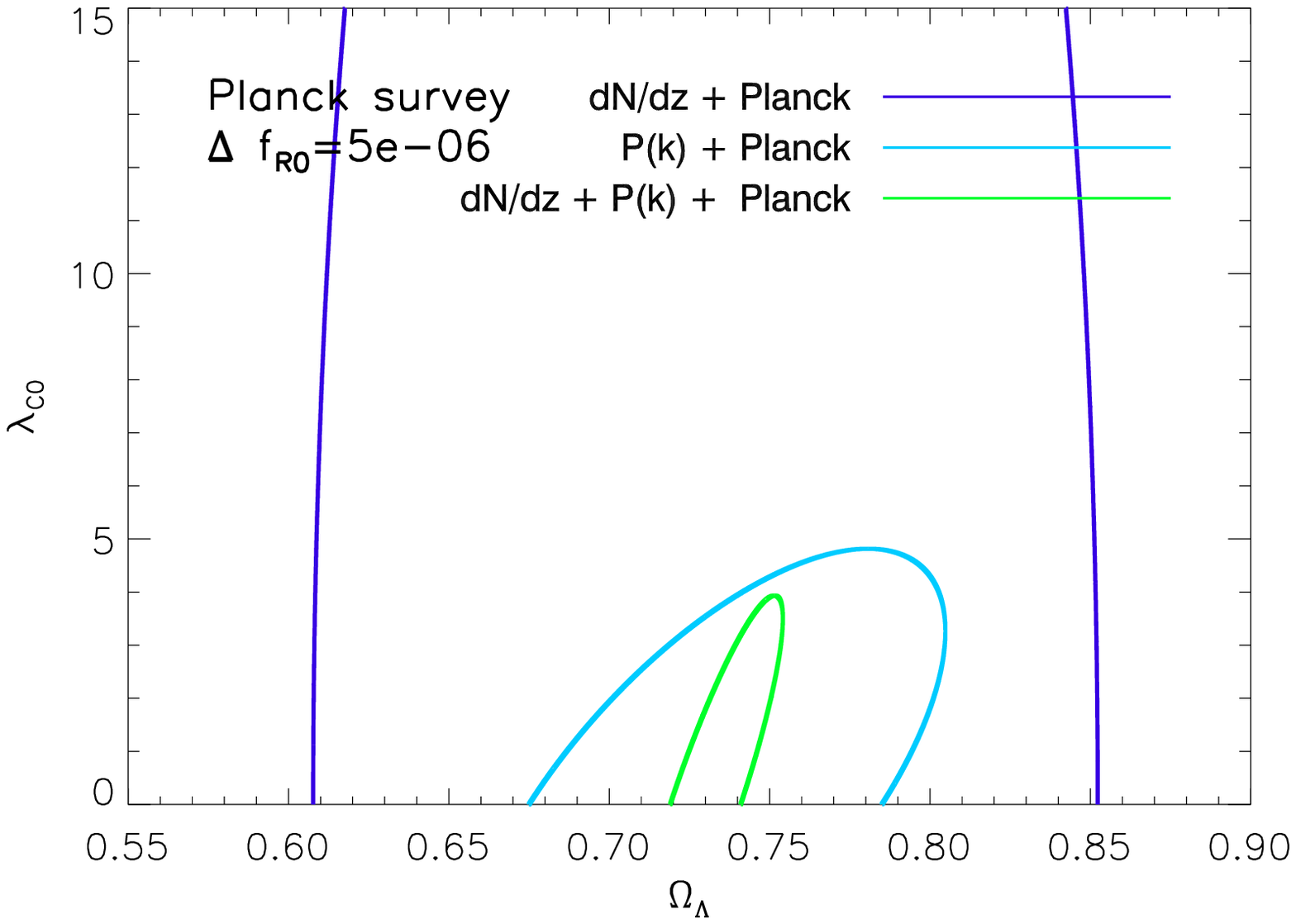}
\includegraphics[width=0.48\textwidth]{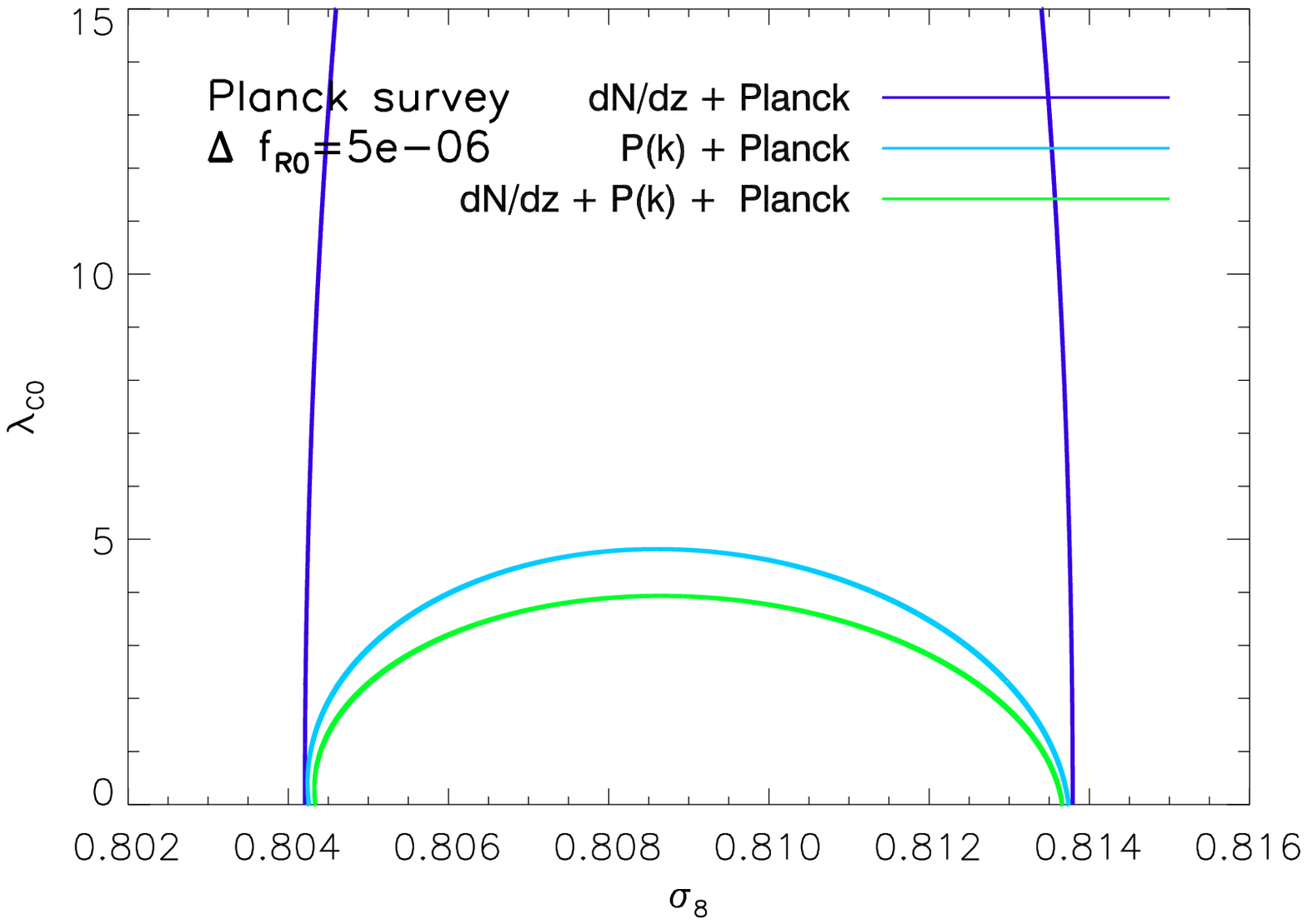}
\includegraphics[width=0.48\textwidth]{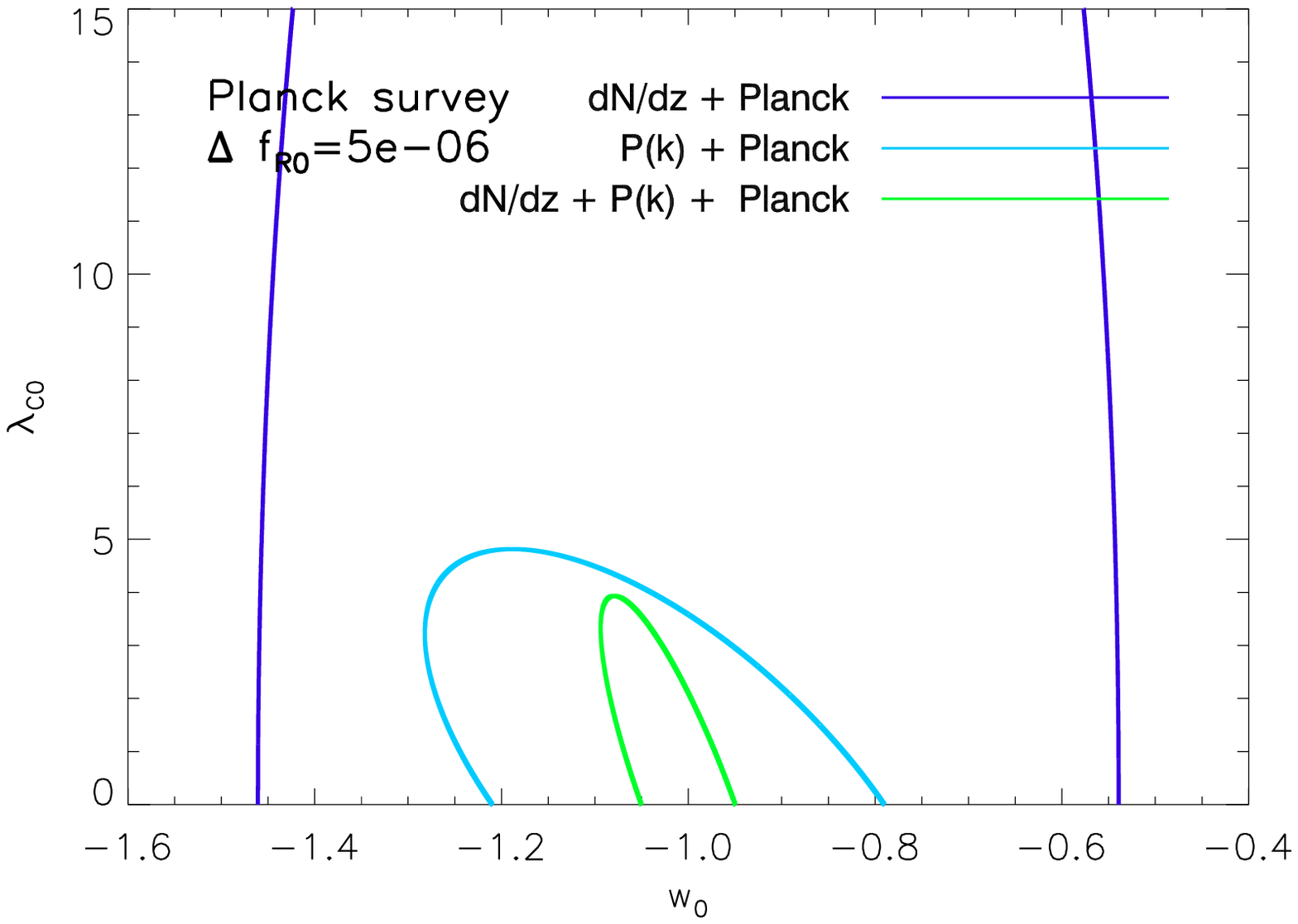}
\includegraphics[width=0.48\textwidth]{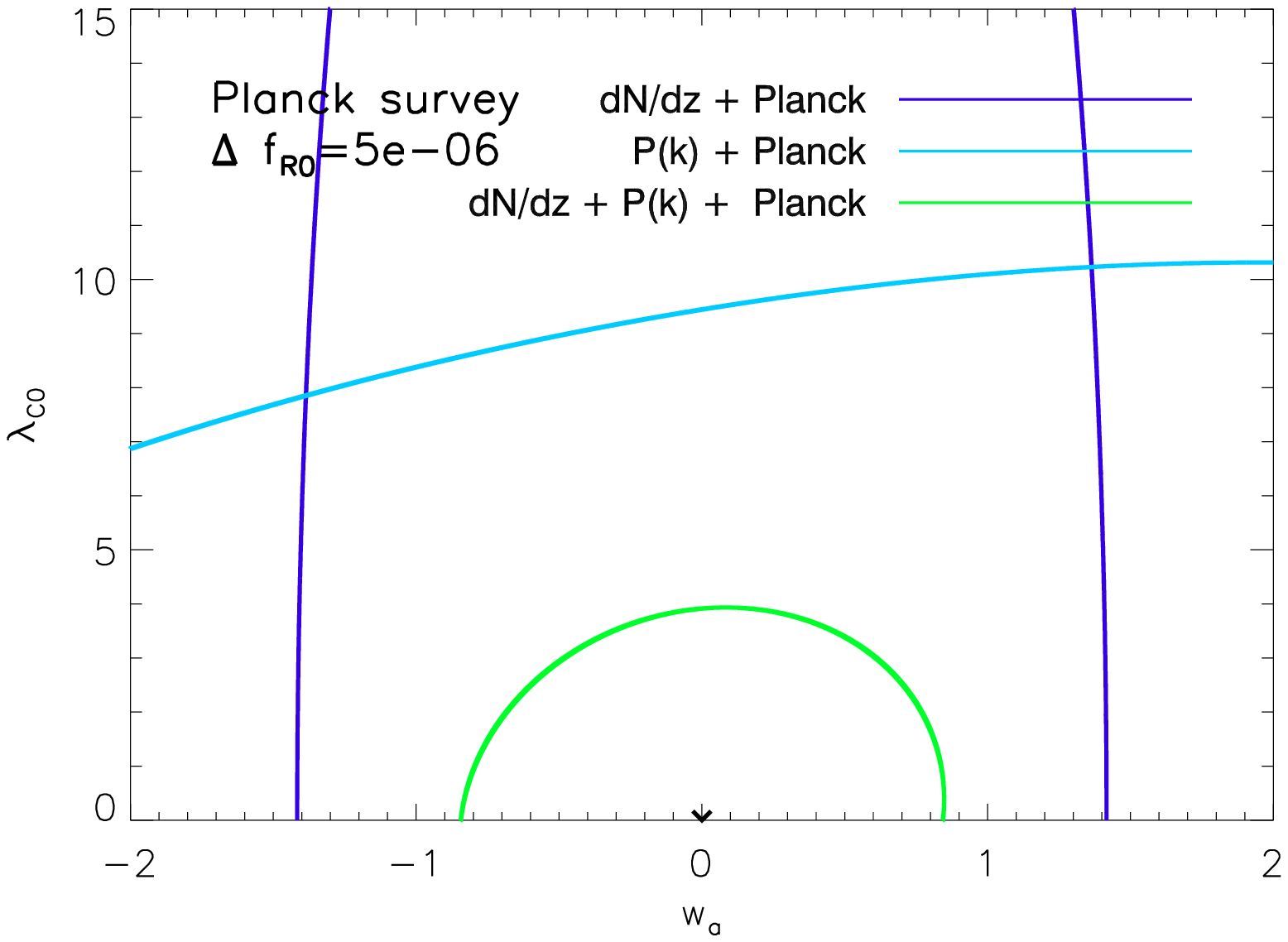}
       \caption{Joint constraints on the Compton wavelength $\lambda_{C0}$ (in Mpc; see \refeq{lamc}) and {\it (counterclockwise from top left)} $\Omega_\Lambda$, $\sigma_8$, $w_o$, and $w_a$.  All curves denote 68\% confidence level, and are for number counts only (blue), power spectrum only (cyan), and combination of the two (green).  
The results are shown for the \planck\ survey with $\Delta f_{R0} = 5\times10^{-6}$. }
     \label{fig:deg}
  \end{center}
\end{figure*}

\reffig{deg} illustrates the most important degeneracies of  $\lambda_{C0}$ with standard  cosmological 
parameters  for the \planck\ survey. Here, we show $\lambda_{C0}$
instead of $f_{R0}$ for purposes of presentation.   
The most prominent degeneracies are with the amount and equation of state of dark energy ($\Omega_\Lambda$, $w_o$ and $w_a$). 
Clearly, the combination of both
observables yields a significant reduction in degeneracies in all cases.  
The degeneracy with dark energy parameters also explains why the combined 
constraints on $f_{R0}$ are slightly better for SPT than for ACTPol, even
though the constraints from number counts and clustering separately are
very similar for the two surveys.  By probing higher redshifts more 
effectively, SPT is able to better break degeneracies with dark energy 
parameters.

Constraints on modified gravity show little with the power spectrum normalization (see \reffig{deg}).  
This is  due to the fact that the high number of clusters detected allows for good characterization of the shape of the mass function beyond its overall normalization.   
Similar but somewhat weaker degeneracies are present for the other surveys.  

\subsection{Uncertainties in scatter of mass observable relations}
\label{sec:priors}

Throughout this work, we have assumed a functional form for the scaling 
relations and then allowed the data to calibrate the parameters that 
characterize it.  
This procedure is possible thanks to the large number of clusters that are 
expected to be detected in these surveys. 
Current strategies for deriving constraints from cluster surveys, however, rely on the calibration of scaling relations as obtained by a small subset of well studied clusters.
In general, allowing more freedom to the scaling relation parameters may avoid biases induced by incorrect scaling relations but can also result in a degradation of the final result.
 In order to investigate the degradation of $\sigma_{f_{R0}}$ due to this self-calibration, 
we repeat the forecasts assuming
different priors on the four ``nuisance'' parameters.  The result is summarized
in \reftab{fom} for the number counts, clustering, and combined, and 
for the four surveys. 

Here, the ``weak prior'' case assumes priors on
the nuisance parameters of $\Delta B_{M,0}=0.05$ and $\Delta\alpha=1$, 
as well as $\Delta\sigma_{M,0}=0.1$ and $\Delta\beta=1$, as suggested by 
comparison between X-ray and lensing cluster mass measurement 
(e.g., the XMM-Newton measurements presented in \cite{Zhang2010}).  The  
combined constraints on $f_{R0}$ are smaller than those for the default, 
no prior case,  
by about 25\%  for Planck, 80\% for ACTpol, and 50\% 
for  SPT and SPTpol. 
The most prominent improvements are seen in number counts only constraints (e.g. a factor 3.8 for SPT).

The ``strong prior'' case assumes that all four nuisance parameters are fixed 
at their fiducial values.  This assumption, which is anyway not realistic, would lead to improvements of about one order of magnitude with respect to the self-calibration results.

This result suggests that although self-calibration does not in general lead to major degradations
in the constraints,  good prior information on 
normalization and scatter in the mass-observable relation can improve 
constraints considerably in partiuclar for the ACTpol and SPT/SPTpol surveys.  

On the other hand, 
it is important to keep in mind that self-calibration relies on a 
specific parametrization of the mass-observable relation and its scatter,
and external measurements are important to validate these assumptions.  
As a worst-case scenario,
we also considered the case of a single mass bin for each survey, i.e.
neglecting all mass information on individual clusters.  The
fully marginalized, combined constraints on $f_{R0}$ (without any priors
on bias and scatter) worsen by approximately a factor of four for Planck and ACTpol.
On the other hand, both SPT and SPTpol constraints degrade by only a factor of three,
since both surveys has a large lever arm in redshift. 
While these constraints are considerably worse than when using mass bins,
the Planck and SPTpol (?) constraints with a single mass bin still improve 
over current upper limits.

\begin{table}
\caption{Relative improvement in constraints on $f_{R0}$, i.e. $\sigma_{f_{R0}}^{\rm no~prior}/\sigma_{f_{R0}}^{\rm weak}$, when including weak priors on the mass-observable relation (see text).  In each case,
$\Delta f_{R0}$ is that given in \reftab{plancksigma} for the corresponding survey/probe.
}
\begin{center}
\begin{tabular}{l cccccc}
\hline\hline
Survey & \multicolumn{3}{c}{$f_{R0} (10^{-6})$} & \multicolumn{3}{c}{$\lambda_{C0}$ (Mpc/h)}  \\
\hline
             & ~$dN/dz$~ & $P(k)$ & ~$dN/dz$~ & ~$dN/dz$~ & $P(k)$ & ~$dN/dz$~ \\
             &  & & $+P(k)$ &  &  & $+P(k)$ \\
\hline

& \multicolumn{6}{c}{weak}\\
\hline
  Planck &     1.00 &      1.02 &      1.23 &      1.02 &      1.01 &      1.11\\ 
     ACTpol &     2.14 &      1.95 &      1.82 &      1.01 &      1.17 &      1.36\\ 
     SPT &     3.80 &      1.02 &      1.48 &      1.03 &      1.01 &      1.21\\ 
  SPTpol &     1.29 &      1.02 &      1.45 &      1.13 &      1.00 &      1.22\\ 
\hline

\hline
\end{tabular} 
\end{center}
\label{t:fom}
\end{table}

\section{Discussion}
\label{sec:disc}

It is worth comparing our forecasted constraints on $f_{R0}$ with
those obtained in \citet{Schmidt:2009am} (\citet{Lombriser:2010mp} obtained
similar constraints).  By combining 49 Chandra X-ray clusters and using geometric constraints from CMB, supernovae, $H_0$, and BAO, they found an 
upper limit of $f_{R0}< 1.4 \times10^{-4}$ (95\% CL), including only the statistical
error.  Our forecasted constraints are tighter by a factor of $\sim 3 - 4$ (ACTpol, SPT, SPTPol) and $\sim 15$ (Planck), respectively.  The main reasons for the 
tighter constraints are: the significantly larger cluster samples yielded by these surveys, the use of the dynamical mass (which improves number count constraints), and
the inclusion of the clustering of clusters as an observable.  
As shown in \refsec{res}, the latter in fact provides the dominant
constraining power for these surveys in the small field limit.  

Furthermore, the constraints in \cite{Schmidt:2009am} are dominated
by the systematic uncertainty in the cluster mass scale, and including
this systematic increases the upper limit to $f_{R0} \lesssim 3\times 10^{-4}$.  
The constraints presented here are marginalized
over the cluster mass scale, and hence already include this systematic.  
Indeed, the combination of power spectrum and number counts is 
essential in order to realize  self-calibration  without loosing  constraining power.

One interesting finding of our study is that the chameleon screening
mechanism, a necessary ingredient in this modified gravity model in
order to satisfy Solar System constraints, has a qualitative impact
on the constraints.  In particular, the number counts by themselves
cannot push constraints below $f_{R0} \sim 10^{-5}$ due to this effect,
while they yield the tightest constraints for larger field values.  
Similarly, the importance of the dynamical mass effect is controlled
by the chameleon threshold.  This is expected to hold for other
modified gravity scenarios as well, as long as the respective
screening mechanism depends mainly on the host halo mass (or potential
well) of the cluster.  On the other hand, screening mechanisms
that mainly depend on the average interior density, such as
the Vainshtein mechanism employed in braneworld and galileon models,
will show a qualitatively different behavior \cite{DGPHM,dynamicalmass}
(see \cite{ClampittEtal} for a study of the related symmetron mechanism).  
For such models, the utility of number counts will not be limited to 
certain parameter ranges.  
Thus, taking into account the screening mechanism is crucial for obtaining 
realistic constraints on any 
viable modified gravity model, both for forecasts and when using actual data.

All of the surveys considered here reach the limit set by the chameleon
mechanism on the constraints from number counts.  The Planck survey
achieves the tightest constraints both due to its large volume,
which reduces the sample variance especially in the cluster power spectrum,
and due to its ability to detect clusters at $z < 0.15$.  For example,
if we limit the Planck cluster sample to $z \geq 0.15$, the
combined constraints in $f_{R0}$ degrade by a factor of four to 
$\sim 2\times 10^{-5}$.  We thus expect that significant improvements
in constraining power are achievable for ground-based SZ surveys 
if the minimum cluster redshift can be reduced.


Several improvements upon our treatment here are possible.  First,
our model for the $f(R)$ effects on mass function and bias of halos
is conservative.  In order to investigate this, we repeated the forecast
using the standard as opposed to modified spherical collapse parameters
in the model prediction \cite{halopaper}.  
In case of the Planck survey, the
fully marginalized, combined constraint is tightened by a factor of $5-6$,
constraining $f_{R0}$ to less than $10^{-6}$.   
This prescription overestimates the $f(R)$ effects in the small
field regime ($f_{R0} \lesssim 10^{-5}$) and thus leads to overly optimistic
constraints.  Nevertheless, the improvement in constraints signals
that it is worth developing a more accurate model for the $f(R)$ effects
on halo mass function and bias (e.g., along the lines of \cite{LiHu}).  
Given the importance of the cluster power spectrum in the constraints,
an accurate model for the modified halo bias will be crucial.  
Furthermore, a model for the cluster power spectrum on mildly
non-linear scales would also lead to tighter constraints by allowing
$k_{\rm max}$ to be increased above the value of $0.1 h/\rm Mpc$ adopted here.

\section{Conclusions}
\label{sec:conc}

The large cluster samples expected from current and upcoming SZ surveys
can be exploited to place tight constraints on modifications to gravity.  
We have shown that the Planck cluster sample will allow for 
more than one order of magnitude improvement in constraints on the field parameter
$f_{R0}$ over current observational constraints,
even when marginalizing over the expansion history (parametrized
by $w_0, w_a$) and bias and scatter in the mass-observable relation.   
Similarly, SPT, SPTPol and ACTPol should provide improvements of about  a factor 3--4.
Using number counts only, the Planck cluster catalog 
should be able to reduce errors to $\sigma_{f_{R0}}=2\times10^{-5}$
in the near future.  
The inclusion of the cluster power spectrum as a probe greatly improves 
results especially in the small field limit.  
The best constraint we obtain is for Planck  (combined constraints, $\sigma_{f_{R0}}=5\times10^{-6}$) and is mainly driven
by the power spectrum.  These constraints push into the regime not ruled out 
by Solar System tests \cite{HuSaw07a}.  
Even with self-calibration, a good understanding of the cluster selection function
will be necessary to realize this potential however.  On the theoretical side,
a better description of the modified gravity effects on halo mass function
and bias should allow for further improvements.  
In addition, the use of a proper likelihood function would 
constitute an important validation of the results obtained here 
with the Fisher matrix approximation.

\acknowledgments
EP and NM acknowledge support from NSF grant AST-0649899.
EP and DM were partially  supported by  NASA grant NNX07AH59G. EP  also acknowledges support from  JPL-Planck subcontract 1290790. She  would like to thank the hospitality of the Aspen Center for Physics for hospitality during the preparation of this work.  
FS would like to thank Wayne Hu for helpful discussions.  
FS is supported by the Gordon and Betty Moore foundation at Caltech.

\appendix

\section{Covariance of Cluster Power Spectra}
\label{app:Pk}

In this appendix we derive the Fisher matrix element for the cross- and
auto-power spectra of clusters binned in mass.  Let $P^{mn}(k)$ denote
the cross-power spectrum between mass bins $m$ and $n$.  In this section
we will suppress the explicit redshift-dependence for clarity.  The
variance of the cross-power spectrum measured in a narrow $k$ range is given
by
\begin{align}
\sigma^2\left(P^{mn}(k)\right) = \frac{1}{N_{\rm mod}} \Bigg [ &
\left ( P^{mm}(k) + \frac{1}{n_m}\right)
\left ( P^{nn}(k) + \frac{1}{n_j}\right)\nonumber\\
& +\left( P^{mn}(k) + \frac{\delta^{mn}}{n_m}\right)^2 \Bigg].
\end{align}
Here, $n_m$ denotes the comoving number density of clusters in mass bin $i$, and
the number of modes is given by
\be
N_{\rm mod} = \frac{1}{2}\frac{V\:k^2 \Delta k}{2\pi^2},
\ee
where the factor of $1/2$ in front accounts for the fact that the density
field is real, reducing the number of independent modes by one half.  
This factor is sometimes neglected in the literature.
The volume is given by
\be
V(z) = \Omega_s \chi^2(z) \frac{c}{H(z)} \Delta z.  
\ee
Using this, we can derive the general power spectrum Fisher matrix
as
\begin{align}
F_{\alpha\beta}=\frac{1}{4\pi^2}\sum_{i,j}\sum_{l,m} & \Bigg [ \frac{\partial \ln P^{mn}(k_m,z_l)}{\partial p_\alpha} \frac{\partial \ln P^{mn}(k_m,z_l)}{\partial p_\beta}  \nonumber\\
&\; \times V^{mn,\rm eff}(k_m, z_l) k_m^2 \Delta k  \Bigg ],
\end{align}
with
\begin{align}
& \frac{V^{mn,\rm eff}(k_i, z_l)}{V_0(z_l)} = P^{mn}(k_i, z_l)]^2 n_m(z_l) n_n(z_l) \vs
& \times \Big[(n_m P^{mm} + 1)(n_n P^{nn} + 1) + n_m n_n (P^{nm} + \delta^{nm} n_m^{-1})^2\Big]^{-1},\nonumber
\end{align}
where all quantities in the denominator are evaluated at $k_m$ and $z_l$.  
This is \refeq{veff}.

\bibliography{ms}

\begin{thebibliography}{58}
\expandafter\ifx\csname natexlab\endcsname\relax\def\natexlab#1{#1}\fi
\expandafter\ifx\csname bibnamefont\endcsname\relax
  \def\bibnamefont#1{#1}\fi
\expandafter\ifx\csname bibfnamefont\endcsname\relax
  \def\bibfnamefont#1{#1}\fi
\expandafter\ifx\csname citenamefont\endcsname\relax
  \def\citenamefont#1{#1}\fi
\expandafter\ifx\csname url\endcsname\relax
  \def\url#1{\texttt{#1}}\fi
\expandafter\ifx\csname urlprefix\endcsname\relax\def\urlprefix{URL }\fi
\providecommand{\bibinfo}[2]{#2}
\providecommand{\eprint}[2][]{\url{#2}}

\bibitem[{\citenamefont{{Tinker} et~al.}(2008)\citenamefont{{Tinker},
  {Kravtsov}, {Klypin}, {Abazajian}, {Warren}, {Yepes}, {Gottlober}, and
  {Holz}}}]{Tin08}
\bibinfo{author}{\bibfnamefont{J.~L.} \bibnamefont{{Tinker}}},
  \bibinfo{author}{\bibfnamefont{A.~V.} \bibnamefont{{Kravtsov}}},
  \bibinfo{author}{\bibfnamefont{A.}~\bibnamefont{{Klypin}}},
  \bibinfo{author}{\bibfnamefont{K.}~\bibnamefont{{Abazajian}}},
  \bibinfo{author}{\bibfnamefont{M.~S.} \bibnamefont{{Warren}}},
  \bibinfo{author}{\bibfnamefont{G.}~\bibnamefont{{Yepes}}},
  \bibinfo{author}{\bibfnamefont{S.}~\bibnamefont{{Gottlober}}},
  \bibnamefont{and} \bibinfo{author}{\bibfnamefont{D.~E.}
  \bibnamefont{{Holz}}}, \bibinfo{journal}{ArXiv e-prints}
  \textbf{\bibinfo{volume}{803}} (\bibinfo{year}{2008}), \eprint{0803.2706}.

\bibitem[{\citenamefont{{Sheth} and {Tormen}}(1999)}]{SheTor99}
\bibinfo{author}{\bibfnamefont{R.}~\bibnamefont{{Sheth}}} \bibnamefont{and}
  \bibinfo{author}{\bibfnamefont{B.}~\bibnamefont{{Tormen}}},
  \bibinfo{journal}{\mnras} \textbf{\bibinfo{volume}{308}},
  \bibinfo{pages}{119} (\bibinfo{year}{1999}).

\bibitem[{\citenamefont{{Viana} and {Liddle}}(1999)}]{Viana99}
\bibinfo{author}{\bibfnamefont{P.~T.~P.} \bibnamefont{{Viana}}}
  \bibnamefont{and} \bibinfo{author}{\bibfnamefont{A.~R.}
  \bibnamefont{{Liddle}}}, \bibinfo{journal}{\mnras}
  \textbf{\bibinfo{volume}{303}}, \bibinfo{pages}{535} (\bibinfo{year}{1999}).

\bibitem[{\citenamefont{{Pierpaoli} et~al.}(2001)\citenamefont{{Pierpaoli},
  {Scott}, and {White}}}]{Pierpaoli01}
\bibinfo{author}{\bibfnamefont{E.}~\bibnamefont{{Pierpaoli}}},
  \bibinfo{author}{\bibfnamefont{D.}~\bibnamefont{{Scott}}}, \bibnamefont{and}
  \bibinfo{author}{\bibfnamefont{M.}~\bibnamefont{{White}}},
  \bibinfo{journal}{\mnras} \textbf{\bibinfo{volume}{325}}, \bibinfo{pages}{77}
  (\bibinfo{year}{2001}), \eprint{arXiv:astro-ph/0010039}.

\bibitem[{\citenamefont{{Pierpaoli} et~al.}(2003)\citenamefont{{Pierpaoli},
  {Borgani}, {Scott}, and {White}}}]{Pierpaoli03}
\bibinfo{author}{\bibfnamefont{E.}~\bibnamefont{{Pierpaoli}}},
  \bibinfo{author}{\bibfnamefont{S.}~\bibnamefont{{Borgani}}},
  \bibinfo{author}{\bibfnamefont{D.}~\bibnamefont{{Scott}}}, \bibnamefont{and}
  \bibinfo{author}{\bibfnamefont{M.}~\bibnamefont{{White}}},
  \bibinfo{journal}{\mnras} \textbf{\bibinfo{volume}{342}},
  \bibinfo{pages}{163} (\bibinfo{year}{2003}), \eprint{arXiv:astro-ph/0210567}.

\bibitem[{\citenamefont{{Borgani} et~al.}(2001)\citenamefont{{Borgani},
  {Rosati}, {Tozzi}, {Stanford}, {Eisenhardt}, {Lidman}, {Holden}, {Della
  Ceca}, {Norman}, and {Squires}}}]{Borgani01}
\bibinfo{author}{\bibfnamefont{S.}~\bibnamefont{{Borgani}}},
  \bibinfo{author}{\bibfnamefont{P.}~\bibnamefont{{Rosati}}},
  \bibinfo{author}{\bibfnamefont{P.}~\bibnamefont{{Tozzi}}},
  \bibinfo{author}{\bibfnamefont{S.~A.} \bibnamefont{{Stanford}}},
  \bibinfo{author}{\bibfnamefont{P.~R.} \bibnamefont{{Eisenhardt}}},
  \bibinfo{author}{\bibfnamefont{C.}~\bibnamefont{{Lidman}}},
  \bibinfo{author}{\bibfnamefont{B.}~\bibnamefont{{Holden}}},
  \bibinfo{author}{\bibfnamefont{R.}~\bibnamefont{{Della Ceca}}},
  \bibinfo{author}{\bibfnamefont{C.}~\bibnamefont{{Norman}}}, \bibnamefont{and}
  \bibinfo{author}{\bibfnamefont{G.}~\bibnamefont{{Squires}}},
  \bibinfo{journal}{\apj} \textbf{\bibinfo{volume}{561}}, \bibinfo{pages}{13}
  (\bibinfo{year}{2001}), \eprint{arXiv:astro-ph/0106428}.

\bibitem[{\citenamefont{{Allen}
  et~al.}(2003{\natexlab{a}})\citenamefont{{Allen}, {Schmidt}, {Fabian}, and
  {Ebeling}}}]{Allen03a}
\bibinfo{author}{\bibfnamefont{S.~W.} \bibnamefont{{Allen}}},
  \bibinfo{author}{\bibfnamefont{R.~W.} \bibnamefont{{Schmidt}}},
  \bibinfo{author}{\bibfnamefont{A.~C.} \bibnamefont{{Fabian}}},
  \bibnamefont{and}
  \bibinfo{author}{\bibfnamefont{H.}~\bibnamefont{{Ebeling}}},
  \bibinfo{journal}{\mnras} \textbf{\bibinfo{volume}{342}},
  \bibinfo{pages}{287} (\bibinfo{year}{2003}{\natexlab{a}}),
  \eprint{arXiv:astro-ph/0208394}.

\bibitem[{\citenamefont{{Pierpaoli}}(2004)}]{Pierpaoli04}
\bibinfo{author}{\bibfnamefont{E.}~\bibnamefont{{Pierpaoli}}}, in
  \emph{\bibinfo{booktitle}{Astrophysics and Space Science Library}}, edited by
  \bibinfo{editor}{\bibnamefont{{M.~Plionis}}} (\bibinfo{year}{2004}), vol.
  \bibinfo{volume}{301} of \emph{\bibinfo{series}{Astrophysics and Space
  Science Library}}, pp. \bibinfo{pages}{93--+}.

\bibitem[{\citenamefont{{Allen}
  et~al.}(2003{\natexlab{b}})\citenamefont{{Allen}, {Schmidt}, and
  {Bridle}}}]{Allen03b}
\bibinfo{author}{\bibfnamefont{S.~W.} \bibnamefont{{Allen}}},
  \bibinfo{author}{\bibfnamefont{R.~W.} \bibnamefont{{Schmidt}}},
  \bibnamefont{and} \bibinfo{author}{\bibfnamefont{S.~L.}
  \bibnamefont{{Bridle}}}, \bibinfo{journal}{\mnras}
  \textbf{\bibinfo{volume}{346}}, \bibinfo{pages}{593}
  (\bibinfo{year}{2003}{\natexlab{b}}), \eprint{arXiv:astro-ph/0306386}.

\bibitem[{\citenamefont{{Vikhlinin} et~al.}(2009)\citenamefont{{Vikhlinin},
  {Kravtsov}, {Burenin}, {Ebeling}, {Forman}, {Hornstrup}, {Jones}, {Murray},
  {Nagai}, {Quintana} et~al.}}]{Vikhlinin09}
\bibinfo{author}{\bibfnamefont{A.}~\bibnamefont{{Vikhlinin}}},
  \bibinfo{author}{\bibfnamefont{A.~V.} \bibnamefont{{Kravtsov}}},
  \bibinfo{author}{\bibfnamefont{R.~A.} \bibnamefont{{Burenin}}},
  \bibinfo{author}{\bibfnamefont{H.}~\bibnamefont{{Ebeling}}},
  \bibinfo{author}{\bibfnamefont{W.~R.} \bibnamefont{{Forman}}},
  \bibinfo{author}{\bibfnamefont{A.}~\bibnamefont{{Hornstrup}}},
  \bibinfo{author}{\bibfnamefont{C.}~\bibnamefont{{Jones}}},
  \bibinfo{author}{\bibfnamefont{S.~S.} \bibnamefont{{Murray}}},
  \bibinfo{author}{\bibfnamefont{D.}~\bibnamefont{{Nagai}}},
  \bibinfo{author}{\bibfnamefont{H.}~\bibnamefont{{Quintana}}},
  \bibnamefont{et~al.}, \bibinfo{journal}{\apj} \textbf{\bibinfo{volume}{692}},
  \bibinfo{pages}{1060} (\bibinfo{year}{2009}), \eprint{0812.2720}.

\bibitem[{\citenamefont{{Martino} et~al.}(2009)\citenamefont{{Martino},
  {Stabenau}, and {Sheth}}}]{Martino09}
\bibinfo{author}{\bibfnamefont{M.~C.} \bibnamefont{{Martino}}},
  \bibinfo{author}{\bibfnamefont{H.~F.} \bibnamefont{{Stabenau}}},
  \bibnamefont{and} \bibinfo{author}{\bibfnamefont{R.~K.}
  \bibnamefont{{Sheth}}}, \bibinfo{journal}{\prd}
  \textbf{\bibinfo{volume}{79}}, \bibinfo{pages}{084013}
  (\bibinfo{year}{2009}), \eprint{0812.0200}.

\bibitem[{\citenamefont{{Diaferio} and {Ostorero}}(2009)}]{Diaferio09}
\bibinfo{author}{\bibfnamefont{A.}~\bibnamefont{{Diaferio}}} \bibnamefont{and}
  \bibinfo{author}{\bibfnamefont{L.}~\bibnamefont{{Ostorero}}},
  \bibinfo{journal}{\mnras} \textbf{\bibinfo{volume}{393}},
  \bibinfo{pages}{215} (\bibinfo{year}{2009}), \eprint{0808.3707}.

\bibitem[{\citenamefont{{Jain} and {Khoury}}(2010)}]{JainKhoury}
\bibinfo{author}{\bibfnamefont{B.}~\bibnamefont{{Jain}}} \bibnamefont{and}
  \bibinfo{author}{\bibfnamefont{J.}~\bibnamefont{{Khoury}}},
  \bibinfo{journal}{Annals of Physics} \textbf{\bibinfo{volume}{325}},
  \bibinfo{pages}{1479} (\bibinfo{year}{2010}), \eprint{1004.3294}.

\bibitem[{\citenamefont{{Clifton} et~al.}(2011)\citenamefont{{Clifton},
  {Ferreira}, {Padilla}, and {Skordis}}}]{CliftonEtal}
\bibinfo{author}{\bibfnamefont{T.}~\bibnamefont{{Clifton}}},
  \bibinfo{author}{\bibfnamefont{P.~G.} \bibnamefont{{Ferreira}}},
  \bibinfo{author}{\bibfnamefont{A.}~\bibnamefont{{Padilla}}},
  \bibnamefont{and}
  \bibinfo{author}{\bibfnamefont{C.}~\bibnamefont{{Skordis}}},
  \bibinfo{journal}{ArXiv e-prints}  (\bibinfo{year}{2011}),
  \eprint{1106.2476}.

\bibitem[{\citenamefont{Schmidt
  et~al.}(2009{\natexlab{a}})\citenamefont{Schmidt, Vikhlinin, and
  Hu}}]{Schmidt:2009am}
\bibinfo{author}{\bibfnamefont{F.}~\bibnamefont{Schmidt}},
  \bibinfo{author}{\bibfnamefont{A.}~\bibnamefont{Vikhlinin}},
  \bibnamefont{and} \bibinfo{author}{\bibfnamefont{W.}~\bibnamefont{Hu}},
  \bibinfo{journal}{Phys. Rev.} \textbf{\bibinfo{volume}{D80}},
  \bibinfo{pages}{083505} (\bibinfo{year}{2009}{\natexlab{a}}),
  \eprint{0908.2457}.

\bibitem[{\citenamefont{Lombriser et~al.}(2010)\citenamefont{Lombriser, Slosar,
  Seljak, and Hu}}]{Lombriser:2010mp}
\bibinfo{author}{\bibfnamefont{L.}~\bibnamefont{Lombriser}},
  \bibinfo{author}{\bibfnamefont{A.}~\bibnamefont{Slosar}},
  \bibinfo{author}{\bibfnamefont{U.}~\bibnamefont{Seljak}}, \bibnamefont{and}
  \bibinfo{author}{\bibfnamefont{W.}~\bibnamefont{Hu}} (\bibinfo{year}{2010}),
  \eprint{1003.3009}.

\bibitem[{\citenamefont{{Rapetti} et~al.}(2010)\citenamefont{{Rapetti},
  {Allen}, {Mantz}, and {Ebeling}}}]{RapettiEtal}
\bibinfo{author}{\bibfnamefont{D.}~\bibnamefont{{Rapetti}}},
  \bibinfo{author}{\bibfnamefont{S.~W.} \bibnamefont{{Allen}}},
  \bibinfo{author}{\bibfnamefont{A.}~\bibnamefont{{Mantz}}}, \bibnamefont{and}
  \bibinfo{author}{\bibfnamefont{H.}~\bibnamefont{{Ebeling}}},
  \bibinfo{journal}{\mnras} \textbf{\bibinfo{volume}{406}},
  \bibinfo{pages}{1796} (\bibinfo{year}{2010}), \eprint{0911.1787}.

\bibitem[{\citenamefont{{Hu} and {Sawicki}}(2007)}]{HuSaw07a}
\bibinfo{author}{\bibfnamefont{W.}~\bibnamefont{{Hu}}} \bibnamefont{and}
  \bibinfo{author}{\bibfnamefont{I.}~\bibnamefont{{Sawicki}}},
  \bibinfo{journal}{\prd} \textbf{\bibinfo{volume}{76}},
  \bibinfo{pages}{064004} (\bibinfo{year}{2007}), \eprint{arXiv:0705.1158}.

\bibitem[{\citenamefont{Oyaizu}(2008)}]{oyaizu08b}
\bibinfo{author}{\bibfnamefont{H.}~\bibnamefont{Oyaizu}},
  \bibinfo{journal}{Phys. Rev.} \textbf{\bibinfo{volume}{D78}},
  \bibinfo{pages}{123523} (\bibinfo{year}{2008}), \eprint{0807.2449}.

\bibitem[{\citenamefont{Oyaizu et~al.}(2008)\citenamefont{Oyaizu, Lima, and
  Hu}}]{Pkpaper}
\bibinfo{author}{\bibfnamefont{H.}~\bibnamefont{Oyaizu}},
  \bibinfo{author}{\bibfnamefont{M.}~\bibnamefont{Lima}}, \bibnamefont{and}
  \bibinfo{author}{\bibfnamefont{W.}~\bibnamefont{Hu}}, \bibinfo{journal}{Phys.
  Rev.} \textbf{\bibinfo{volume}{D78}}, \bibinfo{pages}{123524}
  (\bibinfo{year}{2008}), \eprint{0807.2462}.

\bibitem[{\citenamefont{Schmidt
  et~al.}(2009{\natexlab{b}})\citenamefont{Schmidt, Lima, Oyaizu, and
  Hu}}]{halopaper}
\bibinfo{author}{\bibfnamefont{F.}~\bibnamefont{Schmidt}},
  \bibinfo{author}{\bibfnamefont{M.~V.} \bibnamefont{Lima}},
  \bibinfo{author}{\bibfnamefont{H.}~\bibnamefont{Oyaizu}}, \bibnamefont{and}
  \bibinfo{author}{\bibfnamefont{W.}~\bibnamefont{Hu}}, \bibinfo{journal}{Phys.
  Rev.} \textbf{\bibinfo{volume}{D79}}, \bibinfo{pages}{083518}
  (\bibinfo{year}{2009}{\natexlab{b}}), \eprint{0812.0545}.

\bibitem[{\citenamefont{{Zhao} et~al.}(2011)\citenamefont{{Zhao}, {Li}, and
  {Koyama}}}]{LiEtal}
\bibinfo{author}{\bibfnamefont{G.-B.} \bibnamefont{{Zhao}}},
  \bibinfo{author}{\bibfnamefont{B.}~\bibnamefont{{Li}}}, \bibnamefont{and}
  \bibinfo{author}{\bibfnamefont{K.}~\bibnamefont{{Koyama}}},
  \bibinfo{journal}{\prd} \textbf{\bibinfo{volume}{83}}, \bibinfo{eid}{044007}
  (\bibinfo{year}{2011}), \eprint{1011.1257}.

\bibitem[{\citenamefont{{Williamson} et~al.}(2011)\citenamefont{{Williamson},
  {Benson}, {High}, {Vanderlinde}, {Ade}, {Aird}, {Andersson}, {Armstrong},
  {Ashby}, {Bautz} et~al.}}]{Williamson11}
\bibinfo{author}{\bibfnamefont{R.}~\bibnamefont{{Williamson}}},
  \bibinfo{author}{\bibfnamefont{B.~A.} \bibnamefont{{Benson}}},
  \bibinfo{author}{\bibfnamefont{F.~W.} \bibnamefont{{High}}},
  \bibinfo{author}{\bibfnamefont{K.}~\bibnamefont{{Vanderlinde}}},
  \bibinfo{author}{\bibfnamefont{P.~A.~R.} \bibnamefont{{Ade}}},
  \bibinfo{author}{\bibfnamefont{K.~A.} \bibnamefont{{Aird}}},
  \bibinfo{author}{\bibfnamefont{K.}~\bibnamefont{{Andersson}}},
  \bibinfo{author}{\bibfnamefont{R.}~\bibnamefont{{Armstrong}}},
  \bibinfo{author}{\bibfnamefont{M.~L.~N.} \bibnamefont{{Ashby}}},
  \bibinfo{author}{\bibfnamefont{M.}~\bibnamefont{{Bautz}}},
  \bibnamefont{et~al.}, \bibinfo{journal}{\apj} \textbf{\bibinfo{volume}{738}},
  \bibinfo{pages}{139} (\bibinfo{year}{2011}), \eprint{1101.1290}.

\bibitem[{\citenamefont{{Vanderlinde}
  et~al.}(2010{\natexlab{a}})\citenamefont{{Vanderlinde}, {Crawford}, {de
  Haan}, {Dudley}, {Shaw}, {Ade}, {Aird}, {Benson}, {Bleem}, {Brodwin}
  et~al.}}]{Vanderlinde10}
\bibinfo{author}{\bibfnamefont{K.}~\bibnamefont{{Vanderlinde}}},
  \bibinfo{author}{\bibfnamefont{T.~M.} \bibnamefont{{Crawford}}},
  \bibinfo{author}{\bibfnamefont{T.}~\bibnamefont{{de Haan}}},
  \bibinfo{author}{\bibfnamefont{J.~P.} \bibnamefont{{Dudley}}},
  \bibinfo{author}{\bibfnamefont{L.}~\bibnamefont{{Shaw}}},
  \bibinfo{author}{\bibfnamefont{P.~A.~R.} \bibnamefont{{Ade}}},
  \bibinfo{author}{\bibfnamefont{K.~A.} \bibnamefont{{Aird}}},
  \bibinfo{author}{\bibfnamefont{B.~A.} \bibnamefont{{Benson}}},
  \bibinfo{author}{\bibfnamefont{L.~E.} \bibnamefont{{Bleem}}},
  \bibinfo{author}{\bibfnamefont{M.}~\bibnamefont{{Brodwin}}},
  \bibnamefont{et~al.}, \bibinfo{journal}{\apj} \textbf{\bibinfo{volume}{722}},
  \bibinfo{pages}{1180} (\bibinfo{year}{2010}{\natexlab{a}}),
  \eprint{1003.0003}.

\bibitem[{\citenamefont{{Marriage} et~al.}(2011)\citenamefont{{Marriage},
  {Acquaviva}, {Ade}, {Aguirre}, {Amiri}, {Appel}, {Barrientos}, {Battistelli},
  {Bond}, {Brown} et~al.}}]{Marriage11}
\bibinfo{author}{\bibfnamefont{T.~A.} \bibnamefont{{Marriage}}},
  \bibinfo{author}{\bibfnamefont{V.}~\bibnamefont{{Acquaviva}}},
  \bibinfo{author}{\bibfnamefont{P.~A.~R.} \bibnamefont{{Ade}}},
  \bibinfo{author}{\bibfnamefont{P.}~\bibnamefont{{Aguirre}}},
  \bibinfo{author}{\bibfnamefont{M.}~\bibnamefont{{Amiri}}},
  \bibinfo{author}{\bibfnamefont{J.~W.} \bibnamefont{{Appel}}},
  \bibinfo{author}{\bibfnamefont{L.~F.} \bibnamefont{{Barrientos}}},
  \bibinfo{author}{\bibfnamefont{E.~S.} \bibnamefont{{Battistelli}}},
  \bibinfo{author}{\bibfnamefont{J.~R.} \bibnamefont{{Bond}}},
  \bibinfo{author}{\bibfnamefont{B.}~\bibnamefont{{Brown}}},
  \bibnamefont{et~al.}, \bibinfo{journal}{\apj} \textbf{\bibinfo{volume}{737}},
  \bibinfo{pages}{61} (\bibinfo{year}{2011}), \eprint{1010.1065}.

\bibitem[{\citenamefont{{Planck Collaboration}
  et~al.}(2011{\natexlab{a}})\citenamefont{{Planck Collaboration}, {Ade},
  {Aghanim}, {Arnaud}, {Ashdown}, {Aumont}, {Baccigalupi}, {Balbi}, {Banday},
  {Barreiro} et~al.}}]{Planck11a}
\bibinfo{author}{\bibnamefont{{Planck Collaboration}}},
  \bibinfo{author}{\bibfnamefont{P.~A.~R.} \bibnamefont{{Ade}}},
  \bibinfo{author}{\bibfnamefont{N.}~\bibnamefont{{Aghanim}}},
  \bibinfo{author}{\bibfnamefont{M.}~\bibnamefont{{Arnaud}}},
  \bibinfo{author}{\bibfnamefont{M.}~\bibnamefont{{Ashdown}}},
  \bibinfo{author}{\bibfnamefont{J.}~\bibnamefont{{Aumont}}},
  \bibinfo{author}{\bibfnamefont{C.}~\bibnamefont{{Baccigalupi}}},
  \bibinfo{author}{\bibfnamefont{A.}~\bibnamefont{{Balbi}}},
  \bibinfo{author}{\bibfnamefont{A.~J.} \bibnamefont{{Banday}}},
  \bibinfo{author}{\bibfnamefont{R.~B.} \bibnamefont{{Barreiro}}},
  \bibnamefont{et~al.}, \bibinfo{journal}{ArXiv e-prints}
  (\bibinfo{year}{2011}{\natexlab{a}}), \eprint{1101.2024}.

\bibitem[{\citenamefont{{Planck Collaboration}
  et~al.}(2011{\natexlab{b}})\citenamefont{{Planck Collaboration}, {Aghanim},
  {Arnaud}, {Ashdown}, {Atrio-Barandela}, {Aumont}, {Baccigalupi}, {Balbi},
  {Banday}, {Barreiro} et~al.}}]{Planck11b}
\bibinfo{author}{\bibnamefont{{Planck Collaboration}}},
  \bibinfo{author}{\bibfnamefont{N.}~\bibnamefont{{Aghanim}}},
  \bibinfo{author}{\bibfnamefont{M.}~\bibnamefont{{Arnaud}}},
  \bibinfo{author}{\bibfnamefont{M.}~\bibnamefont{{Ashdown}}},
  \bibinfo{author}{\bibfnamefont{F.}~\bibnamefont{{Atrio-Barandela}}},
  \bibinfo{author}{\bibfnamefont{J.}~\bibnamefont{{Aumont}}},
  \bibinfo{author}{\bibfnamefont{C.}~\bibnamefont{{Baccigalupi}}},
  \bibinfo{author}{\bibfnamefont{A.}~\bibnamefont{{Balbi}}},
  \bibinfo{author}{\bibfnamefont{A.~J.} \bibnamefont{{Banday}}},
  \bibinfo{author}{\bibfnamefont{R.~B.} \bibnamefont{{Barreiro}}},
  \bibnamefont{et~al.}, \bibinfo{journal}{ArXiv e-prints}
  (\bibinfo{year}{2011}{\natexlab{b}}), \eprint{1106.1376}.

\bibitem[{\citenamefont{{Birkinshaw}}(1999)}]{Birkinshaw99}
\bibinfo{author}{\bibfnamefont{M.}~\bibnamefont{{Birkinshaw}}},
  \bibinfo{journal}{\physrep} \textbf{\bibinfo{volume}{310}},
  \bibinfo{pages}{97} (\bibinfo{year}{1999}), \eprint{arXiv:astro-ph/9808050}.

\bibitem[{\citenamefont{{Ameglio} et~al.}(2009)\citenamefont{{Ameglio},
  {Borgani}, {Pierpaoli}, {Dolag}, {Ettori}, and {Morandi}}}]{Ameglio2009}
\bibinfo{author}{\bibfnamefont{S.}~\bibnamefont{{Ameglio}}},
  \bibinfo{author}{\bibfnamefont{S.}~\bibnamefont{{Borgani}}},
  \bibinfo{author}{\bibfnamefont{E.}~\bibnamefont{{Pierpaoli}}},
  \bibinfo{author}{\bibfnamefont{K.}~\bibnamefont{{Dolag}}},
  \bibinfo{author}{\bibfnamefont{S.}~\bibnamefont{{Ettori}}}, \bibnamefont{and}
  \bibinfo{author}{\bibfnamefont{A.}~\bibnamefont{{Morandi}}},
  \bibinfo{journal}{\mnras} \textbf{\bibinfo{volume}{394}},
  \bibinfo{pages}{479} (\bibinfo{year}{2009}), \eprint{0811.2199}.

\bibitem[{\citenamefont{{Rasia} et~al.}(2005)\citenamefont{{Rasia}, {Mazzotta},
  {Borgani}, {Moscardini}, {Dolag}, {Tormen}, {Diaferio}, and
  {Murante}}}]{Rasia2005}
\bibinfo{author}{\bibfnamefont{E.}~\bibnamefont{{Rasia}}},
  \bibinfo{author}{\bibfnamefont{P.}~\bibnamefont{{Mazzotta}}},
  \bibinfo{author}{\bibfnamefont{S.}~\bibnamefont{{Borgani}}},
  \bibinfo{author}{\bibfnamefont{L.}~\bibnamefont{{Moscardini}}},
  \bibinfo{author}{\bibfnamefont{K.}~\bibnamefont{{Dolag}}},
  \bibinfo{author}{\bibfnamefont{G.}~\bibnamefont{{Tormen}}},
  \bibinfo{author}{\bibfnamefont{A.}~\bibnamefont{{Diaferio}}},
  \bibnamefont{and}
  \bibinfo{author}{\bibfnamefont{G.}~\bibnamefont{{Murante}}},
  \bibinfo{journal}{\apjl} \textbf{\bibinfo{volume}{618}}, \bibinfo{pages}{L1}
  (\bibinfo{year}{2005}), \eprint{arXiv:astro-ph/0409650}.

\bibitem[{\citenamefont{{Nagai} et~al.}(2007)\citenamefont{{Nagai},
  {Vikhlinin}, and {Kravtsov}}}]{Nagai2007}
\bibinfo{author}{\bibfnamefont{D.}~\bibnamefont{{Nagai}}},
  \bibinfo{author}{\bibfnamefont{A.}~\bibnamefont{{Vikhlinin}}},
  \bibnamefont{and} \bibinfo{author}{\bibfnamefont{A.~V.}
  \bibnamefont{{Kravtsov}}}, \bibinfo{journal}{\apj}
  \textbf{\bibinfo{volume}{655}}, \bibinfo{pages}{98} (\bibinfo{year}{2007}),
  \eprint{arXiv:astro-ph/0609247}.

\bibitem[{\citenamefont{{Piffaretti} and {Valdarnini}}(2008)}]{PifVal2008}
\bibinfo{author}{\bibfnamefont{R.}~\bibnamefont{{Piffaretti}}}
  \bibnamefont{and}
  \bibinfo{author}{\bibfnamefont{R.}~\bibnamefont{{Valdarnini}}},
  \bibinfo{journal}{\aap} \textbf{\bibinfo{volume}{491}}, \bibinfo{pages}{71}
  (\bibinfo{year}{2008}), \eprint{0808.1111}.

\bibitem[{\citenamefont{{Melin} et~al.}(2006)\citenamefont{{Melin}, {Bartlett},
  and {Delabrouille}}}]{Melin2006}
\bibinfo{author}{\bibfnamefont{J.}~\bibnamefont{{Melin}}},
  \bibinfo{author}{\bibfnamefont{J.~G.} \bibnamefont{{Bartlett}}},
  \bibnamefont{and}
  \bibinfo{author}{\bibfnamefont{J.}~\bibnamefont{{Delabrouille}}},
  \bibinfo{journal}{\aap} \textbf{\bibinfo{volume}{459}}, \bibinfo{pages}{341}
  (\bibinfo{year}{2006}), \eprint{arXiv:astro-ph/0602424}.

\bibitem[{\citenamefont{{Sehgal} et~al.}(2007)\citenamefont{{Sehgal}, {Trac},
  {Huffenberger}, and {Bode}}}]{Sehgal2007}
\bibinfo{author}{\bibfnamefont{N.}~\bibnamefont{{Sehgal}}},
  \bibinfo{author}{\bibfnamefont{H.}~\bibnamefont{{Trac}}},
  \bibinfo{author}{\bibfnamefont{K.}~\bibnamefont{{Huffenberger}}},
  \bibnamefont{and} \bibinfo{author}{\bibfnamefont{P.}~\bibnamefont{{Bode}}},
  \bibinfo{journal}{\apj} \textbf{\bibinfo{volume}{664}}, \bibinfo{pages}{149}
  (\bibinfo{year}{2007}), \eprint{arXiv:astro-ph/0612140}.

\bibitem[{\citenamefont{{Malte Sch{\"a}fer} and
  {Bartelmann}}(2007)}]{Schafer2007}
\bibinfo{author}{\bibfnamefont{B.}~\bibnamefont{{Malte Sch{\"a}fer}}}
  \bibnamefont{and}
  \bibinfo{author}{\bibfnamefont{M.}~\bibnamefont{{Bartelmann}}},
  \bibinfo{journal}{\mnras} \textbf{\bibinfo{volume}{377}},
  \bibinfo{pages}{253} (\bibinfo{year}{2007}), \eprint{arXiv:astro-ph/0602406}.

\bibitem[{\citenamefont{{Vanderlinde}
  et~al.}(2010{\natexlab{b}})\citenamefont{{Vanderlinde}, {Crawford}, {de
  Haan}, {Dudley}, {Shaw}, {Ade}, {Aird}, {Benson}, {Bleem}, {Brodwin}
  et~al.}}]{Vanderlinde2010}
\bibinfo{author}{\bibfnamefont{K.}~\bibnamefont{{Vanderlinde}}},
  \bibinfo{author}{\bibfnamefont{T.~M.} \bibnamefont{{Crawford}}},
  \bibinfo{author}{\bibfnamefont{T.}~\bibnamefont{{de Haan}}},
  \bibinfo{author}{\bibfnamefont{J.~P.} \bibnamefont{{Dudley}}},
  \bibinfo{author}{\bibfnamefont{L.}~\bibnamefont{{Shaw}}},
  \bibinfo{author}{\bibfnamefont{P.~A.~R.} \bibnamefont{{Ade}}},
  \bibinfo{author}{\bibfnamefont{K.~A.} \bibnamefont{{Aird}}},
  \bibinfo{author}{\bibfnamefont{B.~A.} \bibnamefont{{Benson}}},
  \bibinfo{author}{\bibfnamefont{L.~E.} \bibnamefont{{Bleem}}},
  \bibinfo{author}{\bibfnamefont{M.}~\bibnamefont{{Brodwin}}},
  \bibnamefont{et~al.}, \bibinfo{journal}{\apj} \textbf{\bibinfo{volume}{722}},
  \bibinfo{pages}{1180} (\bibinfo{year}{2010}{\natexlab{b}}),
  \eprint{1003.0003}.

\bibitem[{\citenamefont{{Niemack} et~al.}(2010)\citenamefont{{Niemack}, {Ade},
  {Aguirre}, {Barrientos}, {Beall}, {Bond}, {Britton}, {Cho}, {Das}, {Devlin}
  et~al.}}]{Niemack2010}
\bibinfo{author}{\bibfnamefont{M.~D.} \bibnamefont{{Niemack}}},
  \bibinfo{author}{\bibfnamefont{P.~A.~R.} \bibnamefont{{Ade}}},
  \bibinfo{author}{\bibfnamefont{J.}~\bibnamefont{{Aguirre}}},
  \bibinfo{author}{\bibfnamefont{F.}~\bibnamefont{{Barrientos}}},
  \bibinfo{author}{\bibfnamefont{J.~A.} \bibnamefont{{Beall}}},
  \bibinfo{author}{\bibfnamefont{J.~R.} \bibnamefont{{Bond}}},
  \bibinfo{author}{\bibfnamefont{J.}~\bibnamefont{{Britton}}},
  \bibinfo{author}{\bibfnamefont{H.~M.} \bibnamefont{{Cho}}},
  \bibinfo{author}{\bibfnamefont{S.}~\bibnamefont{{Das}}},
  \bibinfo{author}{\bibfnamefont{M.~J.} \bibnamefont{{Devlin}}},
  \bibnamefont{et~al.}, in \emph{\bibinfo{booktitle}{Society of Photo-Optical
  Instrumentation Engineers (SPIE) Conference Series}} (\bibinfo{year}{2010}),
  vol. \bibinfo{volume}{7741} of \emph{\bibinfo{series}{Society of
  Photo-Optical Instrumentation Engineers (SPIE) Conference Series}},
  \eprint{1006.5049}.

\bibitem[{\citenamefont{Nojiri and Odintsov}(2006)}]{Nojiri:2006ri}
\bibinfo{author}{\bibfnamefont{S.}~\bibnamefont{Nojiri}} \bibnamefont{and}
  \bibinfo{author}{\bibfnamefont{S.~D.} \bibnamefont{Odintsov}},
  \bibinfo{journal}{Int.J.Geom.Meth.Mod.Phys.} \textbf{\bibinfo{volume}{4}},
  \bibinfo{pages}{06} (\bibinfo{year}{2006}), \eprint{hep-th/0601213}.

\bibitem[{\citenamefont{Sotiriou and Faraoni}(2010)}]{Sotiriou:2008rp}
\bibinfo{author}{\bibfnamefont{T.~P.} \bibnamefont{Sotiriou}} \bibnamefont{and}
  \bibinfo{author}{\bibfnamefont{V.}~\bibnamefont{Faraoni}},
  \bibinfo{journal}{Rev. Mod. Phys.} \textbf{\bibinfo{volume}{82}},
  \bibinfo{pages}{451} (\bibinfo{year}{2010}), \eprint{0805.1726}.

\bibitem[{\citenamefont{Carroll et~al.}(2004)\citenamefont{Carroll, Duvvuri,
  Trodden, and Turner}}]{Caretal03}
\bibinfo{author}{\bibfnamefont{S.~M.} \bibnamefont{Carroll}},
  \bibinfo{author}{\bibfnamefont{V.}~\bibnamefont{Duvvuri}},
  \bibinfo{author}{\bibfnamefont{M.}~\bibnamefont{Trodden}}, \bibnamefont{and}
  \bibinfo{author}{\bibfnamefont{M.~S.} \bibnamefont{Turner}},
  \bibinfo{journal}{Phys. Rev.} \textbf{\bibinfo{volume}{D70}},
  \bibinfo{pages}{043528} (\bibinfo{year}{2004}), \eprint{astro-ph/0306438}.

\bibitem[{\citenamefont{Nojiri and Odintsov}(2003)}]{NojOdi03}
\bibinfo{author}{\bibfnamefont{S.}~\bibnamefont{Nojiri}} \bibnamefont{and}
  \bibinfo{author}{\bibfnamefont{S.~D.} \bibnamefont{Odintsov}},
  \bibinfo{journal}{Phys. Rev.} \textbf{\bibinfo{volume}{D68}},
  \bibinfo{pages}{123512} (\bibinfo{year}{2003}), \eprint{hep-th/0307288}.

\bibitem[{\citenamefont{Capozziello et~al.}(2003)\citenamefont{Capozziello,
  Carloni, and Troisi}}]{Capozziello:2003tk}
\bibinfo{author}{\bibfnamefont{S.}~\bibnamefont{Capozziello}},
  \bibinfo{author}{\bibfnamefont{S.}~\bibnamefont{Carloni}}, \bibnamefont{and}
  \bibinfo{author}{\bibfnamefont{A.}~\bibnamefont{Troisi}},
  \bibinfo{journal}{Recent Res. Dev. Astron. Astrophys.}
  \textbf{\bibinfo{volume}{1}}, \bibinfo{pages}{625} (\bibinfo{year}{2003}),
  \eprint{astro-ph/0303041}.

\bibitem[{\citenamefont{{Khoury} and {Weltman}}(2004)}]{khoury04a}
\bibinfo{author}{\bibfnamefont{J.}~\bibnamefont{{Khoury}}} \bibnamefont{and}
  \bibinfo{author}{\bibfnamefont{A.}~\bibnamefont{{Weltman}}},
  \bibinfo{journal}{\prd} \textbf{\bibinfo{volume}{69}},
  \bibinfo{pages}{044026} (\bibinfo{year}{2004}),
  \eprint{arXiv:astro-ph/0309411}.

\bibitem[{\citenamefont{{Song} et~al.}(2007)\citenamefont{{Song}, {Hu}, and
  {Sawicki}}}]{SonHuSaw07}
\bibinfo{author}{\bibfnamefont{Y.-S.} \bibnamefont{{Song}}},
  \bibinfo{author}{\bibfnamefont{W.}~\bibnamefont{{Hu}}}, \bibnamefont{and}
  \bibinfo{author}{\bibfnamefont{I.}~\bibnamefont{{Sawicki}}},
  \bibinfo{journal}{\prd} \textbf{\bibinfo{volume}{75}},
  \bibinfo{pages}{044004} (\bibinfo{year}{2007}),
  \eprint{arXiv:astro-ph/0610532}.

\bibitem[{\citenamefont{{Ferraro} et~al.}(2011)\citenamefont{{Ferraro},
  {Schmidt}, and {Hu}}}]{FerraroEtal}
\bibinfo{author}{\bibfnamefont{S.}~\bibnamefont{{Ferraro}}},
  \bibinfo{author}{\bibfnamefont{F.}~\bibnamefont{{Schmidt}}},
  \bibnamefont{and} \bibinfo{author}{\bibfnamefont{W.}~\bibnamefont{{Hu}}},
  \bibinfo{journal}{\prd} \textbf{\bibinfo{volume}{83}}, \bibinfo{eid}{063503}
  (\bibinfo{year}{2011}), \eprint{1011.0992}.

\bibitem[{\citenamefont{Navarro et~al.}(1997)\citenamefont{Navarro, Frenk, and
  White}}]{NavFreWhi97}
\bibinfo{author}{\bibfnamefont{J.~F.} \bibnamefont{Navarro}},
  \bibinfo{author}{\bibfnamefont{C.~S.} \bibnamefont{Frenk}}, \bibnamefont{and}
  \bibinfo{author}{\bibfnamefont{S.~D.~M.} \bibnamefont{White}},
  \bibinfo{journal}{Astrophys. J.} \textbf{\bibinfo{volume}{490}},
  \bibinfo{pages}{493} (\bibinfo{year}{1997}), \eprint{astro-ph/9611107}.

\bibitem[{\citenamefont{{Hu} and {Kravtsov}}(2003)}]{HuKravtsov}
\bibinfo{author}{\bibfnamefont{W.}~\bibnamefont{{Hu}}} \bibnamefont{and}
  \bibinfo{author}{\bibfnamefont{A.~V.} \bibnamefont{{Kravtsov}}},
  \bibinfo{journal}{\apj} \textbf{\bibinfo{volume}{584}}, \bibinfo{pages}{702}
  (\bibinfo{year}{2003}), \eprint{arXiv:astro-ph/0203169}.

\bibitem[{\citenamefont{{Bullock} et~al.}(2001)\citenamefont{{Bullock},
  {Kolatt}, {Sigad}, {Somerville}, {Kravtsov}, {Klypin}, {Primack}, and
  {Dekel}}}]{Buletal01}
\bibinfo{author}{\bibfnamefont{J.~S.} \bibnamefont{{Bullock}}},
  \bibinfo{author}{\bibfnamefont{T.~S.} \bibnamefont{{Kolatt}}},
  \bibinfo{author}{\bibfnamefont{Y.}~\bibnamefont{{Sigad}}},
  \bibinfo{author}{\bibfnamefont{R.~S.} \bibnamefont{{Somerville}}},
  \bibinfo{author}{\bibfnamefont{A.~V.} \bibnamefont{{Kravtsov}}},
  \bibinfo{author}{\bibfnamefont{A.~A.} \bibnamefont{{Klypin}}},
  \bibinfo{author}{\bibfnamefont{J.~R.} \bibnamefont{{Primack}}},
  \bibnamefont{and} \bibinfo{author}{\bibfnamefont{A.}~\bibnamefont{{Dekel}}},
  \bibinfo{journal}{\mnras} \textbf{\bibinfo{volume}{321}},
  \bibinfo{pages}{559} (\bibinfo{year}{2001}), \eprint{arXiv:astro-ph/9908159}.

\bibitem[{\citenamefont{{Schmidt}}(2010)}]{dynamicalmass}
\bibinfo{author}{\bibfnamefont{F.}~\bibnamefont{{Schmidt}}},
  \bibinfo{journal}{\prd} \textbf{\bibinfo{volume}{81}},
  \bibinfo{pages}{103002} (\bibinfo{year}{2010}), \eprint{1003.0409}.

\bibitem[{\citenamefont{{Komatsu} et~al.}(2011)\citenamefont{{Komatsu},
  {Smith}, {Dunkley}, {Bennett}, {Gold}, {Hinshaw}, {Jarosik}, {Larson},
  {Nolta}, {Page} et~al.}}]{KomatsuEtal11}
\bibinfo{author}{\bibfnamefont{E.}~\bibnamefont{{Komatsu}}},
  \bibinfo{author}{\bibfnamefont{K.~M.} \bibnamefont{{Smith}}},
  \bibinfo{author}{\bibfnamefont{J.}~\bibnamefont{{Dunkley}}},
  \bibinfo{author}{\bibfnamefont{C.~L.} \bibnamefont{{Bennett}}},
  \bibinfo{author}{\bibfnamefont{B.}~\bibnamefont{{Gold}}},
  \bibinfo{author}{\bibfnamefont{G.}~\bibnamefont{{Hinshaw}}},
  \bibinfo{author}{\bibfnamefont{N.}~\bibnamefont{{Jarosik}}},
  \bibinfo{author}{\bibfnamefont{D.}~\bibnamefont{{Larson}}},
  \bibinfo{author}{\bibfnamefont{M.~R.} \bibnamefont{{Nolta}}},
  \bibinfo{author}{\bibfnamefont{L.}~\bibnamefont{{Page}}},
  \bibnamefont{et~al.}, \bibinfo{journal}{\apjs}
  \textbf{\bibinfo{volume}{192}}, \bibinfo{pages}{18} (\bibinfo{year}{2011}),
  \eprint{1001.4538}.

\bibitem[{\citenamefont{{Lima} and {Hu}}(2005)}]{LimHu05}
\bibinfo{author}{\bibfnamefont{M.}~\bibnamefont{{Lima}}} \bibnamefont{and}
  \bibinfo{author}{\bibfnamefont{W.}~\bibnamefont{{Hu}}},
  \bibinfo{journal}{\prd} \textbf{\bibinfo{volume}{72}},
  \bibinfo{pages}{043006} (\bibinfo{year}{2005}), \eprint{astro-ph/0503363}.

\bibitem[{\citenamefont{{Lewis} et~al.}(2000)\citenamefont{{Lewis},
  {Challinor}, and {Lasenby}}}]{Lewis2000}
\bibinfo{author}{\bibfnamefont{A.}~\bibnamefont{{Lewis}}},
  \bibinfo{author}{\bibfnamefont{A.}~\bibnamefont{{Challinor}}},
  \bibnamefont{and}
  \bibinfo{author}{\bibfnamefont{A.}~\bibnamefont{{Lasenby}}},
  \bibinfo{journal}{\apj} \textbf{\bibinfo{volume}{538}}, \bibinfo{pages}{473}
  (\bibinfo{year}{2000}), \eprint{arXiv:astro-ph/9911177}.

\bibitem[{\citenamefont{{Pritchard} and {Pierpaoli}}(2008)}]{Pritchard2008}
\bibinfo{author}{\bibfnamefont{J.~R.} \bibnamefont{{Pritchard}}}
  \bibnamefont{and}
  \bibinfo{author}{\bibfnamefont{E.}~\bibnamefont{{Pierpaoli}}},
  \bibinfo{journal}{\prd} \textbf{\bibinfo{volume}{78}},
  \bibinfo{pages}{065009} (\bibinfo{year}{2008}), \eprint{0805.1920}.

\bibitem[{\citenamefont{{Albrecht} et~al.}(2006)\citenamefont{{Albrecht},
  {Bernstein}, {Cahn}, {Freedman}, {Hewitt}, {Hu}, {Huth}, {Kamionkowski},
  {Kolb}, {Knox} et~al.}}]{Albrecht2006}
\bibinfo{author}{\bibfnamefont{A.}~\bibnamefont{{Albrecht}}},
  \bibinfo{author}{\bibfnamefont{G.}~\bibnamefont{{Bernstein}}},
  \bibinfo{author}{\bibfnamefont{R.}~\bibnamefont{{Cahn}}},
  \bibinfo{author}{\bibfnamefont{W.~L.} \bibnamefont{{Freedman}}},
  \bibinfo{author}{\bibfnamefont{J.}~\bibnamefont{{Hewitt}}},
  \bibinfo{author}{\bibfnamefont{W.}~\bibnamefont{{Hu}}},
  \bibinfo{author}{\bibfnamefont{J.}~\bibnamefont{{Huth}}},
  \bibinfo{author}{\bibfnamefont{M.}~\bibnamefont{{Kamionkowski}}},
  \bibinfo{author}{\bibfnamefont{E.~W.} \bibnamefont{{Kolb}}},
  \bibinfo{author}{\bibfnamefont{L.}~\bibnamefont{{Knox}}},
  \bibnamefont{et~al.}, \bibinfo{journal}{ArXiv Astrophysics e-prints}
  (\bibinfo{year}{2006}), \eprint{arXiv:astro-ph/0609591}.

\bibitem[{\citenamefont{{Zhang} et~al.}(2010)\citenamefont{{Zhang}, {Okabe},
  {Finoguenov}, {Smith}, {Piffaretti}, {Valdarnini}, {Babul}, {Evrard},
  {Mazzotta}, {Sanderson} et~al.}}]{Zhang2010}
\bibinfo{author}{\bibfnamefont{Y.-Y.} \bibnamefont{{Zhang}}},
  \bibinfo{author}{\bibfnamefont{N.}~\bibnamefont{{Okabe}}},
  \bibinfo{author}{\bibfnamefont{A.}~\bibnamefont{{Finoguenov}}},
  \bibinfo{author}{\bibfnamefont{G.~P.} \bibnamefont{{Smith}}},
  \bibinfo{author}{\bibfnamefont{R.}~\bibnamefont{{Piffaretti}}},
  \bibinfo{author}{\bibfnamefont{R.}~\bibnamefont{{Valdarnini}}},
  \bibinfo{author}{\bibfnamefont{A.}~\bibnamefont{{Babul}}},
  \bibinfo{author}{\bibfnamefont{A.~E.} \bibnamefont{{Evrard}}},
  \bibinfo{author}{\bibfnamefont{P.}~\bibnamefont{{Mazzotta}}},
  \bibinfo{author}{\bibfnamefont{A.~J.~R.} \bibnamefont{{Sanderson}}},
  \bibnamefont{et~al.}, \bibinfo{journal}{\apj} \textbf{\bibinfo{volume}{711}},
  \bibinfo{pages}{1033} (\bibinfo{year}{2010}), \eprint{1001.0780}.

\bibitem[{\citenamefont{{Schmidt} et~al.}(2010)\citenamefont{{Schmidt}, {Hu},
  and {Lima}}}]{DGPHM}
\bibinfo{author}{\bibfnamefont{F.}~\bibnamefont{{Schmidt}}},
  \bibinfo{author}{\bibfnamefont{W.}~\bibnamefont{{Hu}}}, \bibnamefont{and}
  \bibinfo{author}{\bibfnamefont{M.}~\bibnamefont{{Lima}}},
  \bibinfo{journal}{\prd} \textbf{\bibinfo{volume}{81}},
  \bibinfo{pages}{063005} (\bibinfo{year}{2010}), \eprint{0911.5178}.

\bibitem[{\citenamefont{{Clampitt} et~al.}(2011)\citenamefont{{Clampitt},
  {Jain}, and {Khoury}}}]{ClampittEtal}
\bibinfo{author}{\bibfnamefont{J.}~\bibnamefont{{Clampitt}}},
  \bibinfo{author}{\bibfnamefont{B.}~\bibnamefont{{Jain}}}, \bibnamefont{and}
  \bibinfo{author}{\bibfnamefont{J.}~\bibnamefont{{Khoury}}},
  \bibinfo{journal}{ArXiv e-prints}  (\bibinfo{year}{2011}),
  \eprint{1110.2177}.

\bibitem[{\citenamefont{{Li} and {Hu}}(2011)}]{LiHu}
\bibinfo{author}{\bibfnamefont{Y.}~\bibnamefont{{Li}}} \bibnamefont{and}
  \bibinfo{author}{\bibfnamefont{W.}~\bibnamefont{{Hu}}},
  \bibinfo{journal}{ArXiv e-prints}  (\bibinfo{year}{2011}),
  \eprint{1107.5120}.

\end{thebibliography}

 \end{document}